\documentclass[useAMS,usenatbib,usegraphicx]{mn2e}

\usepackage{amssymb,amsfonts,amsmath,amstext,amsgen,amsopn,amsxtra,indentfirst,lscape,times,rotating}

\title[Debris discs around M stars]{Debris discs around M stars: non-existence versus non-detection}
\author[Heng \& Malik]{Kevin Heng $^{1}$\thanks{Email: kevin.heng@csh.unibe.ch (KH)} and Matej Malik$^{2}$\thanks{E-mail: mmalik@student.ethz.ch (MM)}\\
$^{1}$University of Bern, Center for Space and Habitability, Sidlerstrasse 5, CH-3012, Bern, Switzerland\\
$^{2}$ETH Z\"{u}rich, Institute for Astronomy, Wolfgang-Pauli-Strasse 27, CH-8093, Z\"{u}rich, Switzerland}

\begin{document}

\date{Submitted 2012 December 3.  Re-submitted 2013 January 8, 2013 March 25 and 2013 April 4.  Accepted 2013 April 10.}

\pagerange{\pageref{firstpage}--\pageref{lastpage}} \pubyear{2013}

\maketitle

\label{firstpage}

\begin{abstract}
Motivated by the reported dearth of debris discs around M stars, we use survival models to study the occurrence of planetesimal discs around them.  These survival models describe a planetesimal disc with a small number of parameters, determine if it may survive a series of dynamical processes and compute the associated infrared excess.  For the WISE satellite, we demonstrate that the dearth of debris discs around M stars may be attributed to the small semi-major axes generally probed if either: 1. the dust grains behave like blackbodies emitting at a peak wavelength coincident with the observed one; 2. or the grains are hotter than predicted by their blackbody temperatures and emit at peak wavelengths that are shorter than the observed one.  At these small distances from the M star, planetesimals are unlikely to survive or persist for time scales of 300 Myr or longer if the disc is too massive.  Conversely, our survival models allow for the existence of a large population of low-mass debris discs that are too faint to be detected with current instruments.  We gain further confidence in our interpretation by demonstrating the ability to compute infrared excesses for Sun-like stars that are broadly consistent with reported values in the literature.  However, our interpretation becomes less clear and large infrared excesses are allowed if only one of these scenarios holds: 3. the dust grains are hotter than blackbody and predominantly emit at the observed wavelength; 4. or are blackbody in nature and emit at peak wavelengths longer than the observed one.  Both scenarios imply that the parent planetesimals reside at larger distances from the star than inferred if the dust grains behaved like blackbodies.  In all scenarios, we show that the infrared excesses detected at 22 $\mu$m (via WISE) and 70 $\mu$m (via Spitzer) from AU Mic are easily reconciled with its young age (12 Myr).  Conversely, the existence of the old debris disc (2--8 Gyr) from GJ 581 is due to the large semi-major axes probed by the Herschel PACS instrument.  We elucidate the conditions under which stellar wind drag may be neglected when considering dust populations around M stars.  The WISE satellite should be capable of detecting debris discs around young M stars with ages $\sim 10$ Myr.
\end{abstract}


\section{Introduction}
\label{sect:intro}

Recently, \cite{avenhaus12} conducted a search for debris discs in a sample of 85 M stars using data from the Wide-field Infrared Survey Explorer (WISE).  They reported a null detection rate, which they attributed to either the different evolution of dust around M stars or to an age effect.  By contrast, debris disc detections are prevalent around A, F and G stars \citep{rieke05,bryden06,chen06,su06,wyatt07,meyer08,urban12}.  The null result of \cite{avenhaus12} motivates a theoretical interpretation in terms of the \emph{planetesimals} that are widely believed to reside in these discs \citep{krivov08,ht10,kb10}.

The traditional view of debris discs is that they are \emph{dust} discs generated by collisions between parent planetesimals, which are much larger in size \citep{bp93,z01,wyatt08}.  Only \emph{dynamically hot} discs produce debris discs.  These are discs in which collisions occur and produce enough dust such that the infrared emission associated with the dust grains (from reprocessing the incident starlight) exceeds the detection threshold and is hence observable using an infrared telescope.  In other words, debris discs are the progeny of dynamically hot planetesimal discs.  By contrast, \emph{dynamically warm} discs contain planetesimals with smaller sizes and in larger numbers than those found in dynamically hot discs \citep{ht10}.  Collisions occur frequently between the planetesimals, but they are non-destructive and do not produce dust in any significant amount.  However, the planetesimals are small and numerous enough that they provide a non-negligible covering fraction around the star.  It is the planetesimals themselves that reprocess the incident starlight into infrared flux.  In essence, dynamically warm discs may \emph{mimic} debris discs, although most of the debris discs observed so far are not believed to be dynamically warm discs.\footnote{One way of breaking the degeneracy in interpretation is to search for ``silicate features" in the spectral energy distribution of the observed disc.  For example, HD 72095, HD 69830 and $\eta$ Corvi are known to possess such silicate features, which provides direct evidence for the presence of small dust grains (see \citealt{ht10} and references therein).  Another method is to compare the disc size, inferred from fitting a blackbody to the excess emission, with the actual size if the disc is resolved; if the resolved size is larger, then small dust grains are likely to be dominating the emission.  Yet another test is to determine if the far-infrared and sub-millimetre spectrum of the disk follows a blackbody or exhibits a steeper drop-off due to inefficient emission from small grains, which argues for the presence of a dynamically hot disc.}

There are two plausible approaches to modeling dynamically hot and warm discs.  The first approach is to construct a formation model, starting from the birth of the gaseous, dusty, protoplanetary disc and ending with a gas-poor disc populated with planetesimals (and possibly planets).  The multitude of poorly-understood processes associated with planet formation renders this a formidable task.  Parametrizing our ignorance of these processes produces a model with a large number of free parameters.  The second approach is to construct a \emph{survival} model \citep{ht10}, which addresses the following question: given a planetesimal disc described by a small number of parameters, can it survive a series of dynamical processes that act to destroy it on time scales possibly smaller than its age?

\begin{figure}
\begin{center}
\includegraphics[width=\columnwidth]{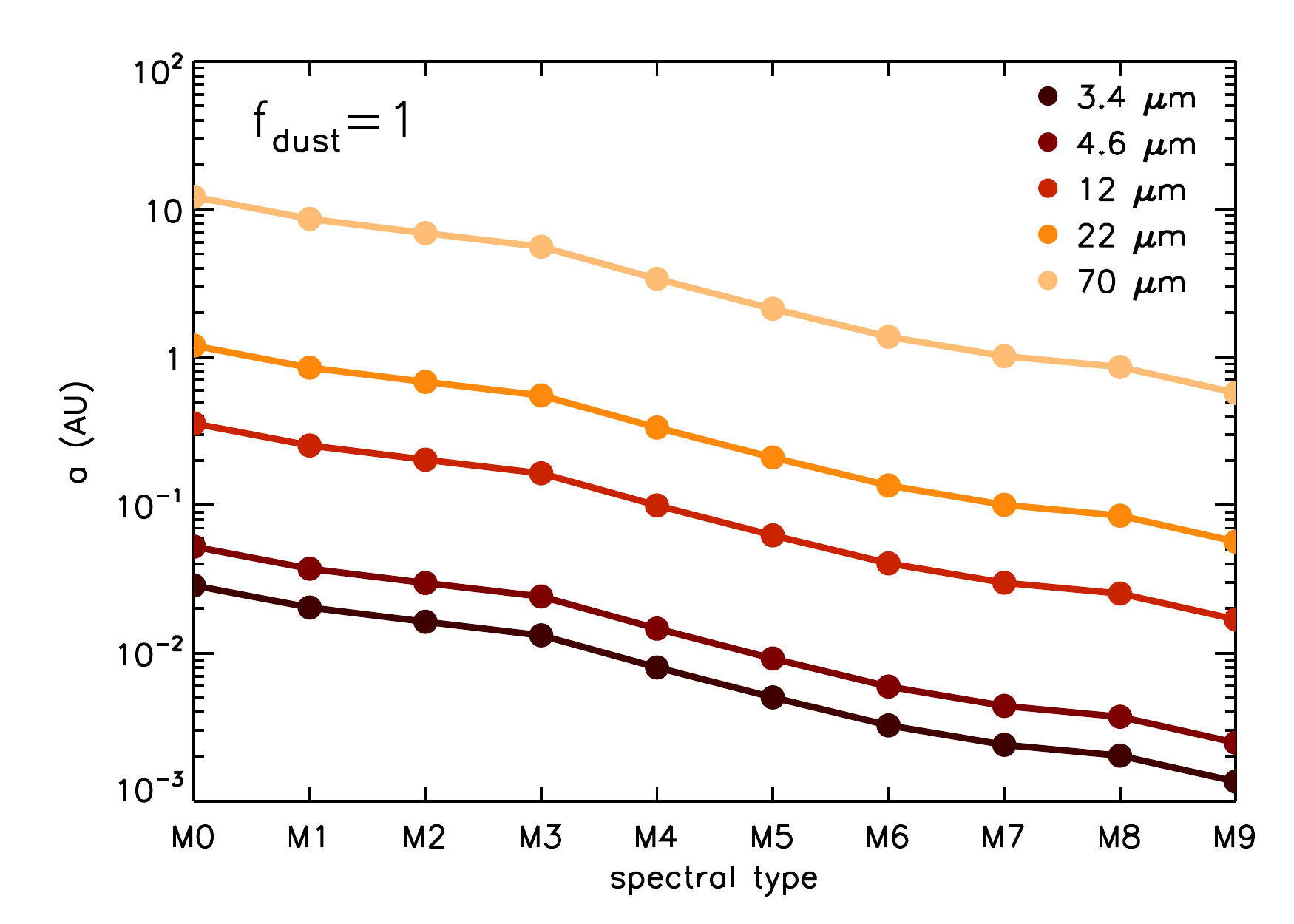}
\end{center}
\vspace{-0.2in}
\caption{Distances from the star probed as a function of the stellar type for the four WISE wavebands, estimated using equation (\ref{eq:adust}).  For comparison, we include the distances associated with the Spitzer 70 $\mu$m channel.  Note that we have assumed $f_{\rm dust}=1$ in these plots.  For small dust grains ($f_{\rm dust}>1$), the distances involved are larger and the chances of survival are more favourable; in this context, these distances are lower limits.  See \S\ref{subsect:ir} for a discussion of other interpretations when $f_{\rm dust} \ne 1$.}
\label{fig:mstar}
\end{figure}

In the present study, we employ the survival models of \cite{ht10} to examine a suite of model discs that are both dynamically hot and warm.  We apply these models to discs around M stars and specialize to the wavebands of WISE.  We demonstrate that a possible reason for the dearth of debris discs observed around M stars is because the WISE wavebands generally probe semi-major axes $\sim 1$ AU or less (Figure \ref{fig:mstar}), where dynamical survival conditions are unfavourable for the parent planetesimals in discs to persist for $\sim 300$ Myr or longer.  This interpretation becomes less clear when the associated dust grains or planetesimals deviate from blackbody behaviour or emit at wavelengths longer than 22 $\mu$m.  We gain further confidence in our interpretation by demonstrating that we can produce infrared excesses for discs around Sun-like stars that are broadly consistent with reported values \citep{bryden06,chen06}, as well as obtain model disc solutions for the infrared excesses observed from AU Mic and GJ 581.

In \S\ref{sect:obs}, we review the observational constraints.  In \S\ref{sect:method}, we construct our survival models of both dynamically hot and warm planetesimal discs.  We present our results in \S\ref{sect:results} and discuss their implications in \S\ref{sect:discussion}.

\section{Observational constraints}
\label{sect:obs}

\cite{avenhaus12} searched for debris discs at 3.4 $\mu$m and 12 $\mu$m around 85 M stars and at 22 $\mu$m around 84 M stars using data from the WISE satellite.  They reported a detection rate of $0.0^{+1.3}_{-0.0}\%$ and estimated the detection threshold for the infrared excess to be $f_{\rm IR,thres} \approx 0.1$--0.3.  We define the infrared excess to be $f_{\rm IR} = (F_{\lambda,{\rm obs}} - F_{\lambda,\star})/F_{\lambda,\star}$ where $F_{\lambda,{\rm obs}}$ is the observed flux at a wavelength $\lambda$ and $F_{\lambda,\star}$ is the photospheric flux from the star at the same wavelength (as given by, e.g., Kurucz models).  While no stellar ages are tabulated, \cite{avenhaus12} do remark that they expect the majority of the stars in their sample to have ages exceeding 300 Myr.  There is little age information and the sample is for nearby M stars.  (See also \citealt{gautier08} and \citealt{simon12}.)

Additionally, \cite{avenhaus12} reported a tentative infrared excess for the relatively bright AU Mic\footnote{Also known as HD 197481 and GJ 803.} (M1 star, $t_\star \approx 12$ Myr), part of the $\beta$ Pic moving group, of $f_{\rm IR} = 0.067 \pm 0.02$ at 22 $\mu$m.  It was previously known that AU Mic has an infrared excess of $f_{\rm IR} = 13.1 \pm 1.5$ at 70 $\mu$m \citep{plavchan09}.

For comparison, we note that $\sim 10\%$ of Sun-like stars, with ages $\sim 10$ Myr to $\sim 1$ Gyr, have reported infrared excesses of $f_{\rm IR} \sim 1$--10 \citep{chen06,bryden06,meyer08}.  For example, \cite{meyer08} find the debris disc frequency around Sun-like stars to be $<4\%$ for $\lambda = 24$ $\mu$m and $t_\star \gtrsim 300$ Myr.  Most systems with detected 24 $\mu$m excesses are inferred to peak at longer wavelengths.  Generally, infrared excesses are more commonly detected at 70 $\mu$m.  (As an aside, we note that \citealt{rizzuto12} detected 22 $\mu$m infrared excesses from B, A and F stars in the 5--17 Myr-old Sco-Cen association with WISE, which they interpreted to emanate from debris discs.)

\section{Method}
\label{sect:method}

Our model contains 9 basic parameters.
\begin{enumerate}

\item \textbf{Star:} the stellar mass $M_\star$, age $t_\star$, effective temperature $T_\star$ and radius $R_\star$.

\item \textbf{Disc:} the total mass $M_{\rm disc}$ and semi-major axis $a$ of the disc.

\item \textbf{Planetesimals:} the radius $r$, radial velocity dispersion $\sigma_r$ and mass density $\rho$ of the planetesimal.

\end{enumerate}
The stellar luminosity is given by ${\cal L}_\star = 4 \pi R_\star^2 \sigma_{\rm SB} T^4_\star$ where $\sigma_{\rm SB}$ is the Stefan-Boltzmann constant.  We obtain the values of the stellar parameters from Table 1 of \cite{kt09}.  We assume $\rho = 3$ g cm$^{-3}$.

\subsection{Survival Models for Dynamically Hot \& Warm Discs}
\label{sect:cond}

In this section, we state the conditions for the survival models of both dynamically hot and warm discs.  Many of these conditions are previously described in \cite{ht10}, but we rewrite them in the form of inequalities involving $r$ and/or $\sigma_r$ and also generalize them for $M_\star \ne M_\odot$.

We examine a disc that is centered at a distance $a$ from its star and extends from $a - f_m a/4$ to $a + f_m a/4$, i.e., it has a width of $f_m a/2$.  When $f_m=1$, we term the disc to extend over an ``octave" following the terminology of \cite{ht10}.

\subsubsection{Common Conditions}

A trivial condition for a disc with a mass $M_{\rm disc}$ is that it needs to contain more than one planetesimal,
\begin{equation}
r < \left( \frac{3 M_{\rm disc}}{4 \pi \rho} \right)^{1/3}.
\end{equation}

We demand that the planetesimal eccentricities and inclinations are not too large: $e_0 < f_e$ where $e_0$ is the root-mean-square (rms) eccentricity and $f_e = 0.5$.  We assume that $i_0/e_0 = 0.5$ where $i_0$ is the rms inclination.  Requiring $e_0 < f_e$ is equivalent to demanding that the disc is thin,
\begin{equation}
\sigma_r < f_e \left( \frac{G M_\star}{2 a} \right)^{1/2}.
\end{equation}

A disc is considered to be dynamically hot or warm when the radial excursions of the planetesimals ($2a e_0$) exceed the typical radial separation between planetesimals,
\begin{equation}
\sigma_r > \frac{\pi f_m \rho r^3}{3 M_{\rm disc}} \left( \frac{G M_\star}{2 a} \right)^{1/2}.
\end{equation}

If the planetesimals are numerous enough that the disc may be approximated as a fluid, then the gravitational stability of a hot disc is described by the usual Toomre criterion \citep{toomre64}.  However, if the fluid approximation breaks down, then the criterion needs to be generalized as described in \S3.2.1 of \cite{ht10}.  Specifically, this occurs when
\begin{equation}
r > \left( \frac{3}{\rho M_\star} \right)^{1/3} \left( \frac{M_{\rm disc}}{f_m} \right)^{2/3}.
\label{eq:gen_toomre}
\end{equation}
When equation (\ref{eq:gen_toomre}) is fulfilled, gravitational instability occurs when
\begin{equation}
\sigma^2_r > \left( \frac{G \rho r^3}{a} \right) \left[ \frac{2}{3} - \frac{\rho r^3 M_\star}{9} \left( \frac{f_m}{M_{\rm disc}} \right)^2 \right].
\end{equation}
Otherwise, the usual Toomre criterion applies,
\begin{equation}
\sigma_r > \frac{M_{\rm disc}}{f_m} \left( \frac{G}{a M_\star} \right)^{1/2}.
\end{equation}

\subsubsection{Dynamically Hot Discs}

A dynamically hot disc necessarily needs to manufacture dust, which requires collisions to occur between its constituent planetesimals.  However, these collisions should not occur too frequently, otherwise the planetesimals will destroy themselves on a time scale that is smaller than the stellar age.  For a planetesimal disc to survive for an age $t_\star$, the collisional time between planetesimals needs to be $t_{\rm coll} > t_\star$.  Such a condition has two regimes demarcated by whether the planetesimals are self-gravitating.  In the regime where gravitational focusing becomes non-negligible, the Safronov number $\Theta$ equals or exceeds unity,
\begin{equation}
r \ge \left( \frac{3}{2 \pi G \rho} \right)^{1/2} \sigma_r.
\label{eq:selfgrav}
\end{equation}
When equation (\ref{eq:selfgrav}) is fulfilled, we apply the following constraint,
\begin{equation}
\sigma_r > \left( \frac{8 f_2 G M_{\rm disc} t_\star r}{\pi f_m} \right)^{1/2} \left( \frac{GM_\star}{a^7} \right)^{1/4},
\end{equation}
where $f_2 \approx 1.521$ (see Appendix A of \citealt{ht10}).  Otherwise, we apply the constraint,
\begin{equation}
r > \frac{12 f_1 M_{\rm disc} t_\star}{\pi^2 f_m \rho} \left( \frac{GM_\star}{a^7} \right)^{1/2},
\end{equation}
where $f_1 \approx 0.690$.

The last condition for the survival of a dynamically hot disc over a time $t_\star$ involves gravitational scattering, which generally produces substantial changes in $e_0$ and $i_0$.  We demand that $t_{\rm grav} > t_\star$, which yields
\begin{equation}
\sigma_r > \left( \frac{\rho M_{\rm disc} t_\star}{3 f_m} \right)^{1/4} G^{5/8} a^{-7/8}  M^{1/8}_\star r^{3/4} \left( S_1 C^\prime \right)^{1/4}.
\end{equation}
The quantity $S_1 = S_1(i_0/e_0)$ generally depends on the ratio of the rms inclination to eccentricity of the planetesimals; we have $S_1(0.5) \approx 4.50$.  Associated with the quantity $C^\prime = C + S_2/S_1$, we have
\begin{equation}
\begin{split}
2 C &= \ln\left( \Lambda^2 + 1 \right) - \ln\left( \Lambda_c^2 + 1 \right) + \frac{1}{\Lambda^2 + 1} - \frac{1}{\Lambda_c^2 + 1}, \\
\Lambda &= \frac{15 \sigma_r^2 a}{8 \pi G \rho r^3} \left[ \sigma_r \left( \frac{a}{G M_\star} \right)^{1/2}  + \left( \frac{8 \pi \rho r^3}{9 M_\star} \right)^{1/3} \right], \\
\Lambda_c &= \frac{15 \sigma_r^2}{4 \pi G \rho r^2} \left\{ 1 + \frac{4 \pi \rho r^2 a}{3 M_\star} \left[ \frac{5 \sigma^2_r a}{2 G M_\star} + \frac{1}{2} \left( \frac{8 \pi \rho r^3}{9 M_\star} \right)^{2/3} \right]^{-1} \right\}^{1/2}. \\
\end{split}
\end{equation}
Furthermore, the quantity $S_2$ contains exponential integrals:
\begin{equation}
\begin{split}
S_2 &\approx 10.13 \left[ W\left( x \right) - W\left( 4x \right) \right], \\
x &= \left( \frac{8 \pi \rho}{9 M_\star} \right)^{2/3} \frac{GM_\star r^2}{2 \sigma^2_r a},\\
W\left(x \right) &\equiv \exp{\left(x\right)} ~E_1\left(x\right),\\
E_1\left(x\right) &\equiv \int^\infty_1 \frac{\exp\left(-x x^\prime \right)}{x^\prime} ~dx^\prime.
\end{split}
\end{equation}

\subsubsection{Dynamically Warm Discs}

In dynamically warm discs, we require that collisions between the planetesimals leave the mass distribution unchanged.  The latter condition is fulfilled when the planetesimals do not form gravitationally bound pairs (i.e., the Safronov number should be less than unity),
\begin{equation}
r < \left( \frac{3}{2 \pi G \rho} \right)^{1/2} \sigma_r.
\end{equation}

The collisions are assumed to be frequent ($t_{\rm coll} < t_\star$),
\begin{equation}
r < \frac{12 f_1 M_{\rm disc} t_\star}{\pi^2 f_m \rho} \left( \frac{GM_\star}{a^7} \right)^{1/2},
\end{equation}
but non-erosive in nature,
\begin{equation}
\sigma_r < \pi \left( \frac{a^7}{GM_\star} \right)^{1/4} \left( \frac{f_m \rho r Q^\ast_{\rm D}}{12 f_1 t_\star M_{\rm disc}} \right)^{1/2}.
\end{equation}
The binding energy per unit mass of the planetesimals is given by
\begin{equation}
Q^\ast_{\rm D} = Q_0 \left[ \left(\frac{r}{r_0}\right)^a + \left(\frac{r}{r_0}\right)^b \right],
\label{eq:bind}
\end{equation}
where $Q_0  \sim 10^4$--$10^7$ erg g$^{-1}$, $r_0 = 2 \times 10^4$ cm, $a=-0.4$ and $b=1.3$ \citep{ba99,sl09}.  We will explore the variation of the normalization $Q_0$ as this carries an uncertainty of several orders of magnitude.

An additional (and weaker) constraint is that the collisions between the planetesimals do not result in significant viscous spreading of the disc,
\begin{equation}
\sigma_r < \frac{\pi}{8} \left( \frac{f_m \rho r}{6 f_1 f_6 t_\star M_{\rm disc}} \right)^{1/2} \left( GM_\star \right)^{1/4} a^{5/4},
\end{equation}
where we have taken the typical energy loss per unit mass in a collision to be $f_6 \sigma^2_r$.  We adopt $f_6=1$.

\subsection{Infrared Emission}
\label{subsect:ir}

\subsubsection{From Dust Grains}

\begin{figure}
\begin{center}
\includegraphics[width=\columnwidth]{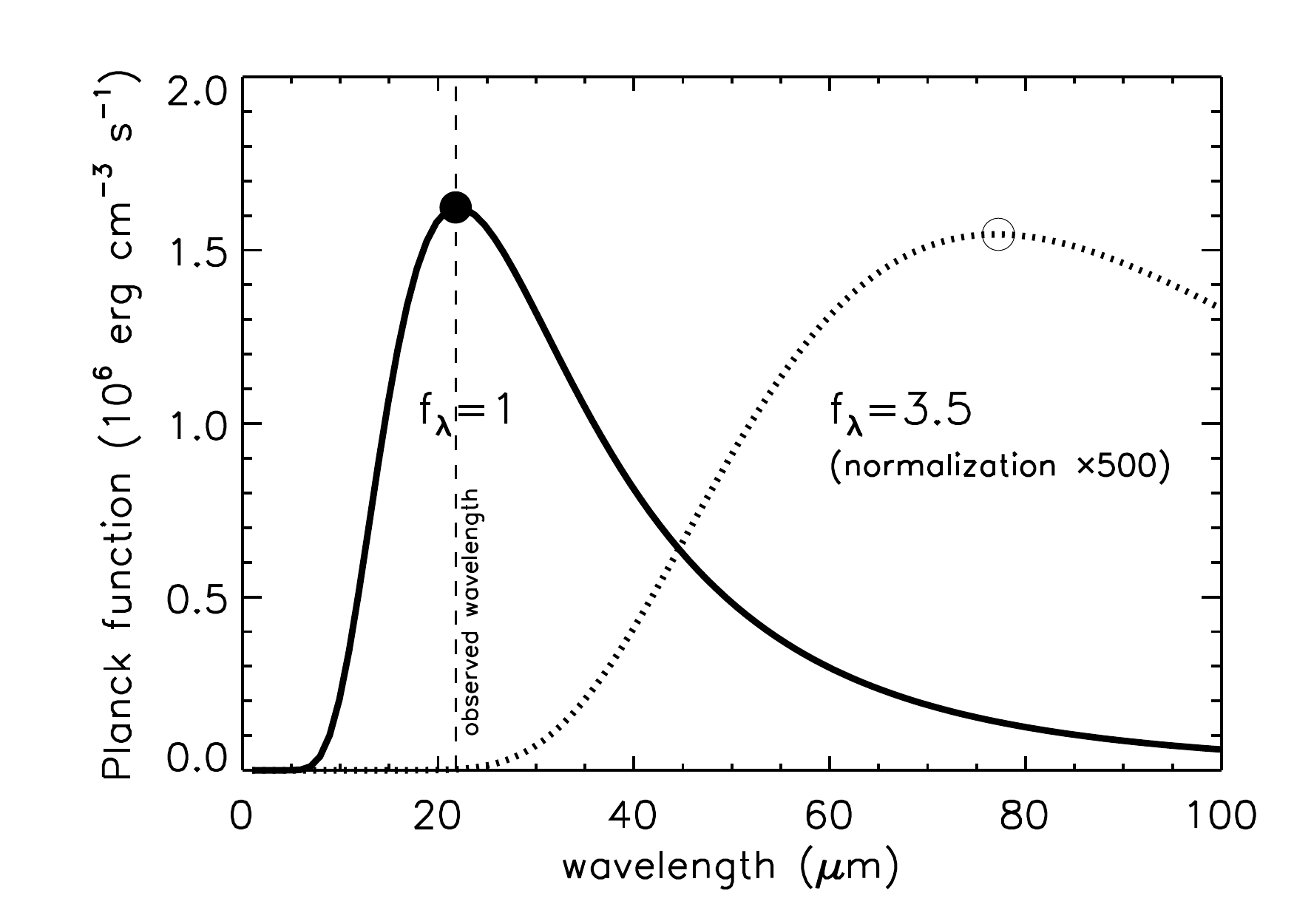}
\end{center}
\vspace{-0.2in}
\caption{Examples of Planck or blackbody functions with $f_\lambda=1$ and $f_\lambda=3.5$, peaking at $\lambda = 22$ and 77 $\mu$m, respectively.}
\label{fig:bbody}
\end{figure}

Consider a dust grain, with a radius $r$, to be located at a distance $a$ from a star.  It is heated to a temperature of
\begin{equation}
T = T_\star \left( \frac{R_\star}{2 a} \right)^{1/2} \left( \frac{Q_{\rm abs}}{Q_{\rm emit}} \right)^{1/4},
\label{eq:bbtemp}
\end{equation}
where $Q_{\rm abs}$ and $Q_{\rm emit}$ are the absorption and emission efficiencies of the grain, respectively.  The preceding equation requires that the spherical grain intersects the incident starlight with a projected cross section of $\pi r^2$ and redistributes the heat over a surface area of $4\pi r^2$.  Following \cite{lestrade12}, we define
\begin{equation}
f_{\rm T} \equiv \left( \frac{Q_{\rm abs}}{Q_{\rm emit}} \right)^{1/4}.
\end{equation}
Blackbody grains have $f_{\rm T}=1$.  Small grains may possess temperatures higher than predicted by their blackbody values ($f_{\rm T} > 1$), because they emit inefficiently at wavelengths larger than their sizes.  For the debris disc around GJ 581, \cite{lestrade12} infer $f_{\rm T} = 3.5^{+0.5}_{-1.0}$.  It is unknown if this is a representative value for the dust populations of debris discs around M stars in general.  We note that for the debris disc around the G star 61 Vir, \cite{wyatt12} find $f_{\rm T}=1.9$.

When grains are re-emitting starlight at a wavelength $\lambda$, an effect to consider is that $\lambda$ may not necessarily be the peak wavelength of the emission profile.  This can be mimicked using the following expression,
\begin{equation}
T = \frac{{\cal C}_{\rm Wien}}{f_\lambda \lambda},
\end{equation}
where ${\cal C}_{\rm Wien} = 2897.7685$ $\mu$m K is Wien's displacement constant and $f_\lambda$ is a dimensionless constant.  Grains with $f_\lambda \ne 1$ emit at a peak wavelength of $f_\lambda \lambda$.  In the example shown in Figure \ref{fig:bbody}, a grain with $f_\lambda=1$ has a temperature of about 132 K and emits at a peak wavelength of 22 $\mu$m.  Also shown is an example with $f_\lambda=3.5$.  In this case, the grain behaves like a blackbody emitting at a peak wavelength of $f_\lambda \lambda \approx 77$ $\mu$m.  Such an approximation allows us to model debris discs that emit at, e.g., $\lambda = 22$ $\mu$m, but do not possess spectral energy distributions that peak at this wavelength.  Note that we are not implying that \cite{lestrade12} inferred $f_\lambda = 3.5$, but rather we have picked $f_\lambda=3.5$ in anticipation of the degeneracy between $f_{\rm T}$ and $f_\lambda$ we will next discuss.

\begin{figure}
\begin{center}
\includegraphics[width=\columnwidth]{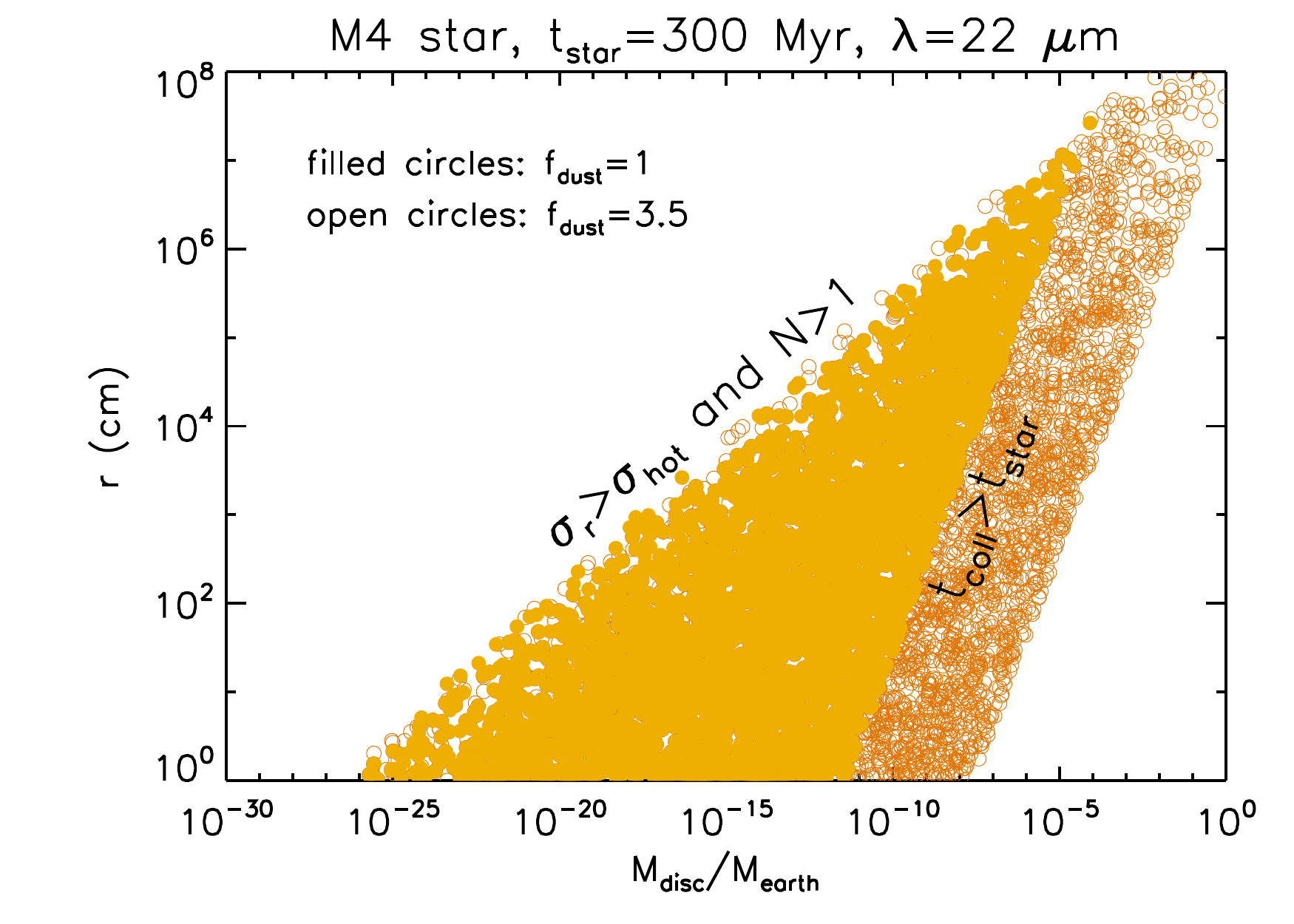}
\includegraphics[width=\columnwidth]{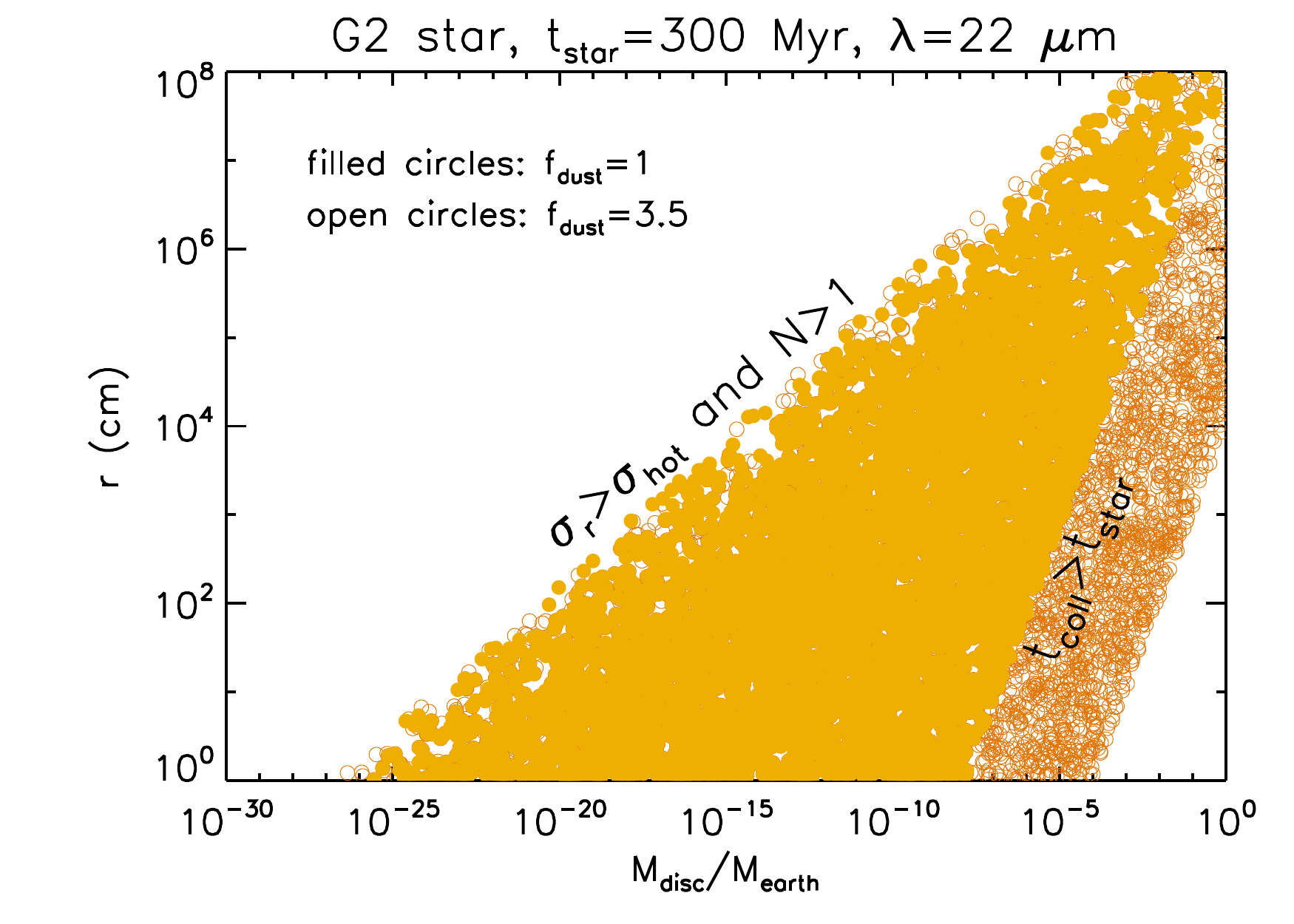}
\end{center}
\vspace{-0.2in}
\caption{Planetesimal radius versus disc mass for a suite of model discs that satisfy the dynamical survival conditions stated in the text.  We examine dynamically hot discs that are capable of generating dust via collisions.  For illustration, we assume $t_\star = 300$ Myr and $\lambda = 22$ $\mu$m.  Top panel: M4 star ($a \approx 0.3$ AU at $\lambda=22$ $\mu$m).  Bottom panel: G2 star ($a \approx 4.5$ AU at $\lambda=22$ $\mu$m).  Note that the labels apply only for the boundaries corresponding to the $f_{\rm dust}=1$ case.}
\label{fig:rmdisc_hot}
\end{figure}

\begin{figure}
\begin{center}
\includegraphics[width=\columnwidth]{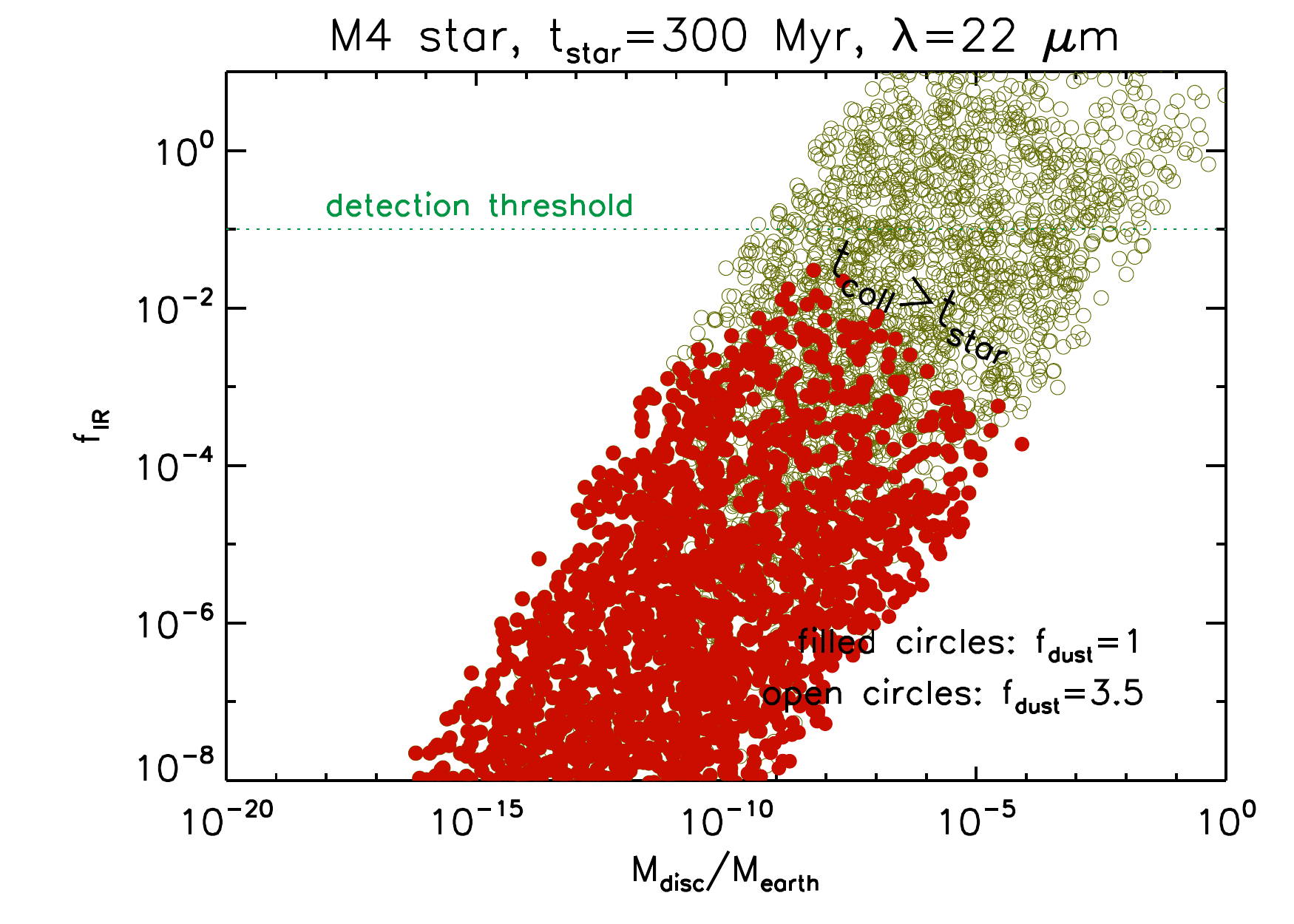}
\includegraphics[width=\columnwidth]{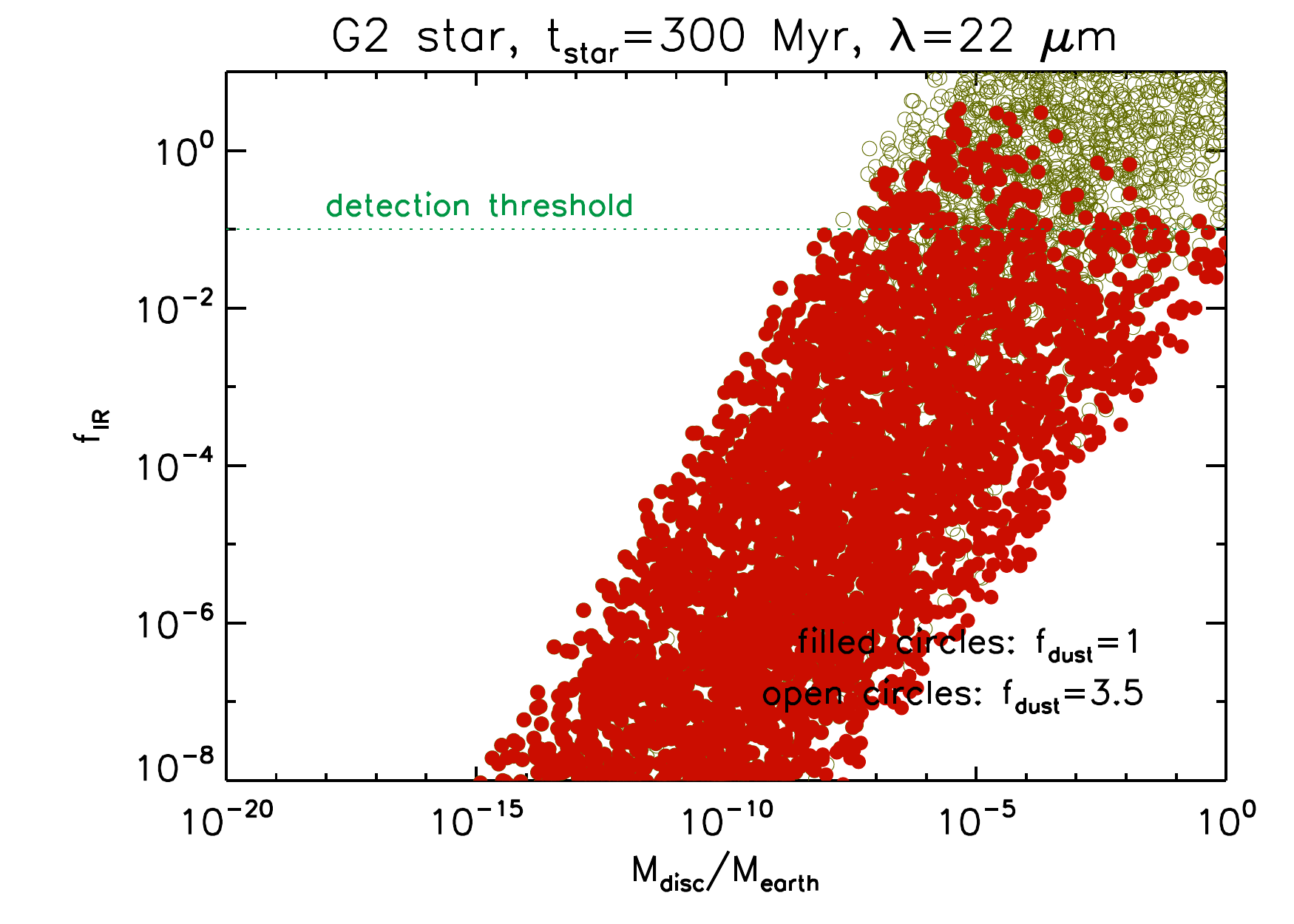}
\end{center}
\vspace{-0.2in}
\caption{Same as Figure \ref{fig:rmdisc_hot} but for the infrared excess at $\lambda=22$ $\mu$m associated with dust produced by colliding planetesimals in dynamically hot discs.  For illustration, the horizontal dotted line shows the detection limit of $f_{\rm IR,thres}=0.1$ associated with the WISE satellite.  Note that the labels apply only for the boundaries corresponding to the $f_{\rm dust}=1$ case.}
\label{fig:fir_hot}
\end{figure}

Given the observed wavelength $\lambda$ as well as the grain and stellar properties, the distance of the grain (and hence that of its parent planetesimals) from the star can be evaluated,
\begin{equation}
a = \frac{R_\star}{2} \left( \frac{T_\star f_{\rm dust} \lambda}{ {\cal C}_{\rm Wien} } \right)^2,
\label{eq:adust}
\end{equation}
where
\begin{equation}
f_{\rm dust} \equiv f_{\rm T} f_\lambda.
\end{equation}
The expression for $f_{\rm dust}$ reflects a number of degeneracies inherent in this parametrization.  For example, a larger value of $a$ may be caused by blackbody ($f_{\rm T}=1$) grains residing at larger distances emitting at a peak wavelength of $f_\lambda \lambda$.  It may also be caused by ``hot" ($f_{\rm T} > 1$) grains with a peak wavelength of emission coincident with the observed wavelength ($f_\lambda=1$).  The ``hotness" of a grain ($f_{\rm T} > 1$) may be offset by it emitting at a shorter peak wavelength ($f_\lambda = 1/f_{\rm T}$), causing it to resemble a blackbody grain emitting at a peak wavelength of $\lambda$.  Strictly speaking, the use of the free parameters $f_{\rm T}$ and $f_\lambda$ are inconsistent with the derivations of equation (\ref{eq:bbtemp}) and Wien's law, which assume a blackbody.  In the absence of broader knowledge on the values of these parameters, we consider both $f_{\rm dust}=1$ and 3.5 in our calculations.  A value of $f_{\rm dust} \ne 1$ may be interpreted in any of the ways just discussed.

Although our planetesimals are assumed to be mono-disperse (i.e., of a single radius/size), we allow collisions between them to produce a collisional cascade of dust grains.  As described in \S7.4 of \cite{ht10}, our calculation of the infrared excess does not require the specification of the values of the minimum and maximum radii in the cascade.  (Physically, a minimum radius is not needed because small grains do not contribute significantly to the infrared emission if $r \ll \lambda/2\pi$.)  The infrared excess at a given wavelength is
\begin{equation}
f_{\rm IR}(\lambda) = \frac{B_\lambda\left( \lambda, T \right)}{B_\lambda\left( \lambda, T_\star \right)} \frac{K}{\left(4 - q\right)\left(q-3\right) R_\star^2} \left( \frac{\lambda}{2\pi} \right)^{3-q}.
\label{eq:fir_dust}
\end{equation}
Unlike for $a$, there is a subtlety concerning the specification of the grain temperature $T$.  It is ${\cal C}_{\rm Wien} f_{\rm T} / \lambda$ if one wishes to model hot (non-blackbody) grains emittingly predominantly at the observed wavelength $\lambda$.  However, it is ${\cal C}_{\rm Wien}/f_\lambda\lambda$ if one wishes to model blackbody grains with lower temperatures being observed at a wavelength $\lambda$ shorter than the peak wavelength of emission ($f_\lambda \lambda$).  Such considerations have no effect on elucidating the allowed region of parameter space for the planetesimals.  The boundaries of the allowed region of $f_{\rm IR}$ versus $M_{\rm disc}$ are largely unchanged with the maximum value of the infrared excess, of the entire ensemble, differing by a factor $\sim 10$ between $f_{\rm T}=3.5$ and $f_\lambda=3.5$.  Instead, we find that the main effect of $f_{\rm dust}$ is in changing the value of $a \propto f^2_{\rm dust}$.  Thus, to keep our analysis (and figures) simple, we use ${\cal C}_{\rm Wien}/\lambda$ as the input grain temperature in equation (\ref{eq:fir_dust}) while allowing for $f_{\rm dust} \ne 1$ in equation (\ref{eq:adust}).

The quantity $K$ is the normalization factor in the radius distribution of the dust grains,
\begin{equation}
K = \left(\frac{45}{8r}\right)^{1/2} \left( \frac{M_{\rm disc}}{\pi \rho} \right) \left( 1 + \frac{2 \pi f_2 G \rho r^2}{3 f_1 \sigma^2_r} \right)^{1/2} \left( \frac{2 Q^\ast_{\rm D,dust}}{\sigma^2_r} \right)^{5/12},
\end{equation}
where $Q^\ast_{\rm D,dust} \approx 10^7$ erg g$^{-1}$ is the binding energy per unit mass of \emph{dust grains} in the collisional cascade and is approximately constant with mass (unlike for the planetesimals).  The Planck function is given by $B_\lambda = (2 h c^2/\lambda^5)/[\exp(hc/\lambda k_{\rm B} T) - 1]$ where $h$ is the Planck constant, $c$ is the speed of light and $k_{\rm B}$ is the Boltzmann constant.  Denoting the number of dust grains by $N_{\rm dust}$, the size distribution of the dust grains takes the form $dN_{\rm dust}/dr = K r^{-q}$ with $q=7/2$.

\subsubsection{From Planetesimals}

The planetesimals are large enough that it is safe to assume $r/\lambda \gg 1$ and $f_{\rm T}=1$.  However, one still needs to consider the possibility that they may emit at a peak wavelength of $f_\lambda \lambda$ where $f_\lambda \ne 1$.  Thus, we set $f_{\rm dust} = f_\lambda$ when using equation (\ref{eq:adust}).  The infrared excess at a given wavelength is 
\begin{equation}
f_{\rm IR} = \frac{3 M_{\rm disc}}{4 \pi \rho r R^2_\star} \frac{B_\lambda\left( \lambda, T \right)}{B_\lambda\left( \lambda, T_\star \right)}.
\end{equation}
Since $f_{\rm dust} = f_\lambda$, the input planetesimal temperature $T$ takes the form of ${\cal C}_{\rm Wien}/f_{\rm dust}\lambda$.

\section{Results}
\label{sect:results}

\subsection{Basic Model}

\begin{figure}
\begin{center}
\includegraphics[width=\columnwidth]{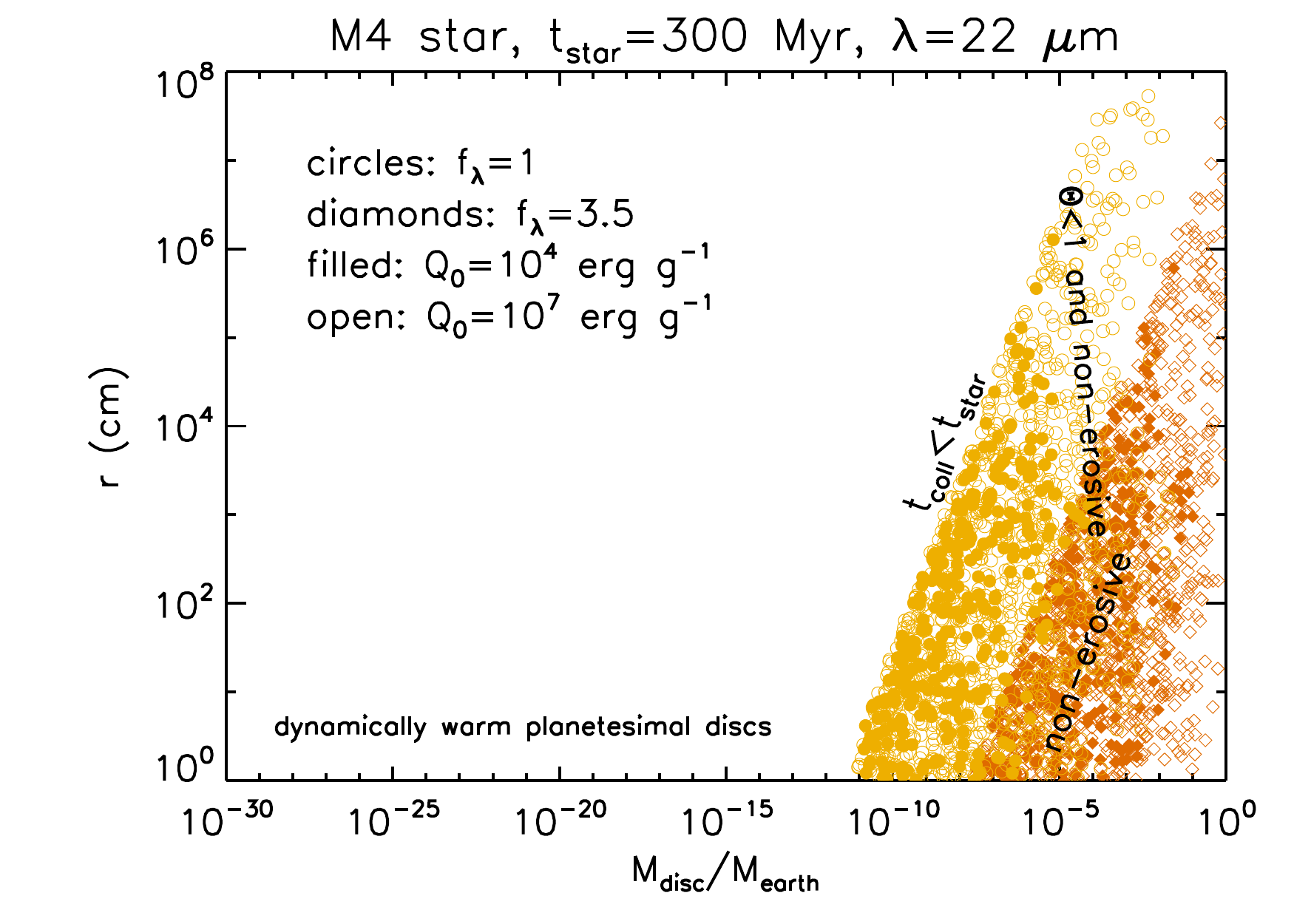}
\includegraphics[width=\columnwidth]{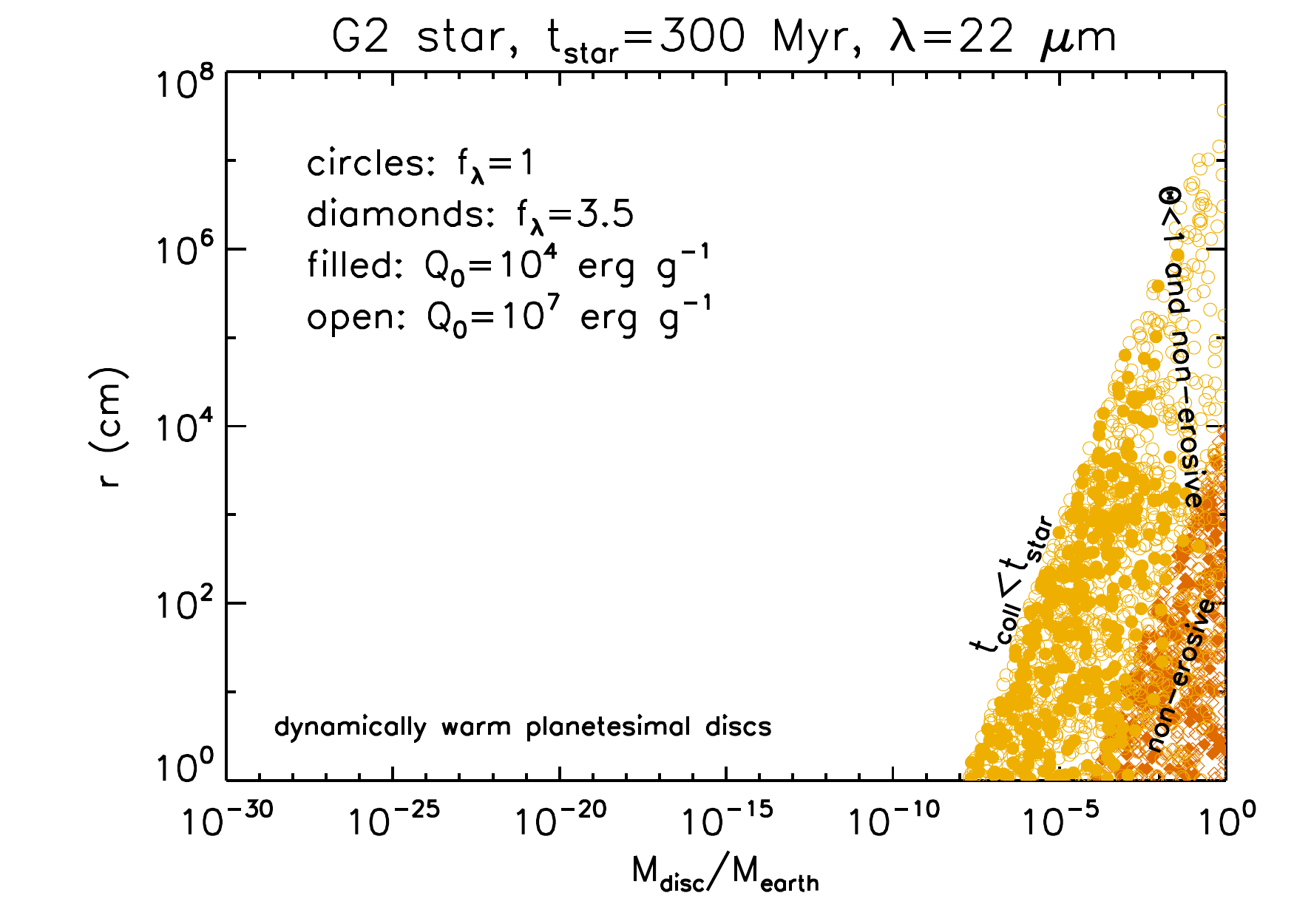}
\end{center}
\vspace{-0.2in}
\caption{Same as Figure \ref{fig:rmdisc_hot} but for dynamically warm planetesimal discs.  Note that the boundaries shown are only for $f_{\rm dust}=1$ and $Q_0=10^4$ erg g$^{-1}$.}
\label{fig:rmdisc_warm}
\end{figure}

\begin{figure}
\begin{center}
\includegraphics[width=\columnwidth]{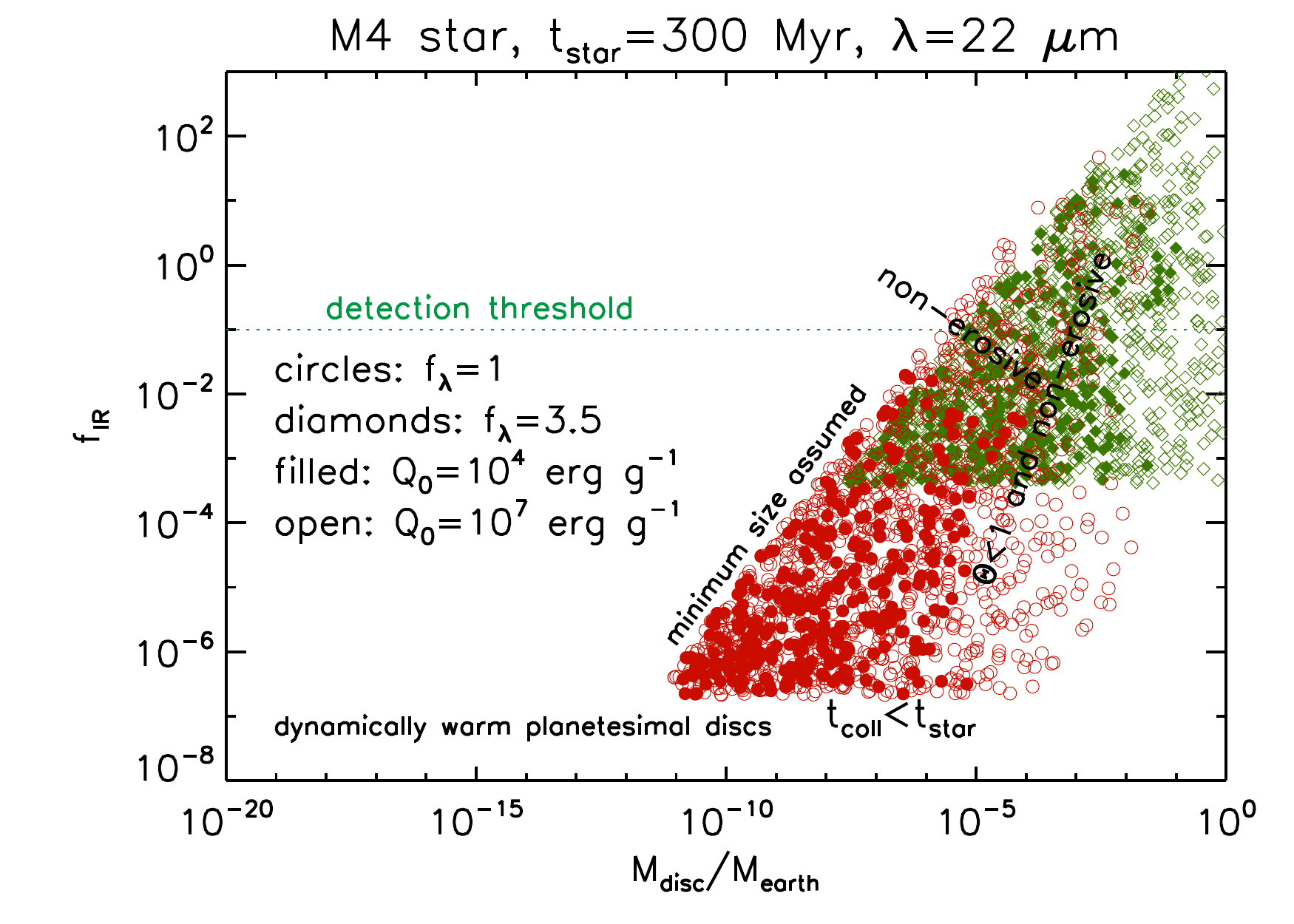}
\includegraphics[width=\columnwidth]{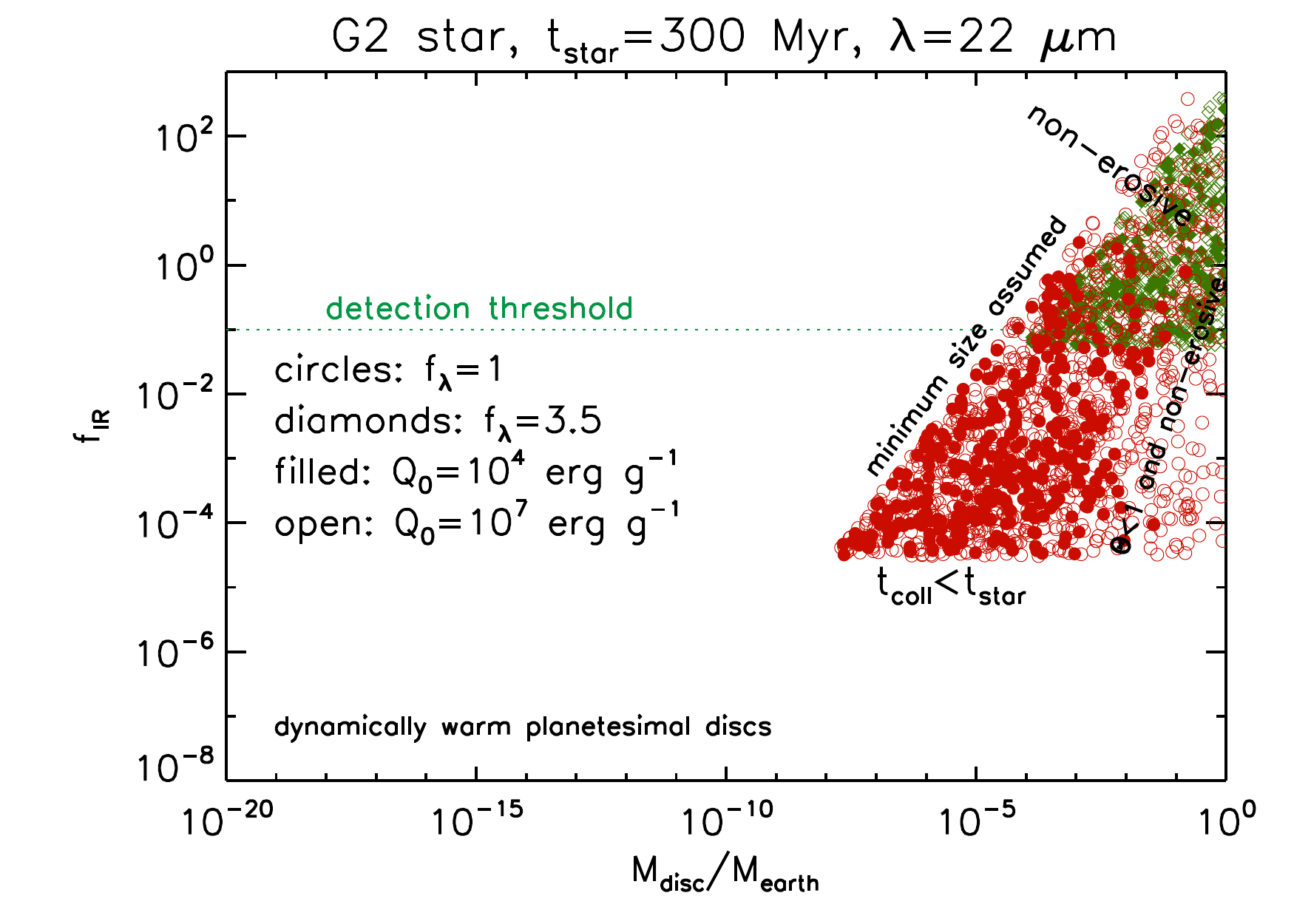}
\end{center}
\vspace{-0.2in}
\caption{Same as Figure \ref{fig:fir_hot} but for dynamically warm planetesimal discs.  For these discs, the infrared emission originates from the planetesimals themselves as secondary dust is not manufactured in significant amounts.  Note that the boundaries shown are only for $f_{\rm dust}=1$ and $Q_0=10^4$ erg g$^{-1}$.}
\label{fig:fir_warm}
\end{figure}

Only three out of the 9 parameters in our model are either unconstrained by the observations or not easily specified from first principles: the disc mass $M_{\rm disc}$, the planetesimal radius $r$ and the radial velocity dispersion of the planetesimal $\sigma_r$.  We use a Monte Carlo method to explore the variation of these parameters over a wide range: $10^{-30} \le M_{\rm disc}/M_\oplus \le 1$, $1 \mbox{ cm} \le r \le 10^4$ km and $1 \mbox{ cm s}^{-1} \le \sigma_r \le 100$ km s$^{-1}$.  The very wide range of parameter values probed ensures that the allowed parameter space is constrained by physics rather than artificially selected boundaries.  We assume that these parameters are uniformly distributed over the stated, logarithmic ranges of values.  For each set of stellar parameters, we generate 30000 model planetesimal discs.  For each disc, we check the list of survival conditions described in \S\ref{sect:cond}; if all of the relevant conditions are fulfilled, we deem the disk to have ``survived".  Its existence---but not necessarily its detection---is thus plausible.

Figure \ref{fig:rmdisc_hot} shows the suite of dynamically hot planetesimal discs that satisfy the dynamical survival conditions in the parameter space of $r$ versus $M_{\rm disc}$.  For illustration, we pick $t_\star=300$ Myr and set the value of $a$ to correspond to $\lambda=22$ $\mu$m as stated in equation (\ref{eq:adust}).  We examine conditions both around a M4 (top panel) and a G2 (bottom panel) star.  As shown in Figure \ref{fig:rmdisc_hot}, the allowed parameter space is severely constrained by the survival conditions.  In particular, the $t_{\rm coll} > t_\star$ constraint sets a lower bound on $r$ at a given $\sigma_r$, while demanding that the discs are dynamically hot ($\sigma_r > \sigma_{\rm hot}$) and contain more than one planetesimal ($N>1$) set an upper bound.  Collectively, these conditions set an upper limit on the maximum planetesimal radius allowed, which is $\sim 100$ km for our 300 Myr-old M4 star if $f_{\rm dust}=1$ and the disc mass is $\sim 10^{-5} M_\oplus$.  The other survival conditions do not dictate the bounds of the allowed region, but rather the density of points within the region.

One may ask if the differences in the allowed parameter regions seen for the M4 and G2 stars are due to the differences in the $M_\star$, $R_\star$ and $T_\star$ values or because of the difference in the value of $a$ at a stated wavelength (equation [\ref{eq:adust}]).  To resolve this question, we fix $a$ and examine plots (not shown) similar to those in Figure \ref{fig:rmdisc_hot} for both the M4 and G2 stars.  We find that the allowed regions of parameter space are both qualitatively and quantitatively similar.  Thus, the most important effect of varying the stellar type is in the determination of $a$ at a fixed $\lambda$, which then affects all of the previously stated survival conditions involving $a$.  In Figure \ref{fig:rmdisc_hot}, the fact that more massive discs are allowed around a G2 star at $\lambda=22$ $\mu$m is primarily an effect of $a \approx 4.5$ AU (compared to $a \approx 0.3$ AU around a M4 star) for $f_{\rm dust}=1$.  Dynamical processes tend to be more forgiving and act more slowly when one is located farther away from the star.  When $f_{\rm dust} = 3.5$, the corresponding values of $a$ are larger and thus more massive planetesimal discs are generally allowed.

At this point, it is worth emphasizing that the focus of our study is neither to examine planetesimal and debris discs around Sun-like stars, nor to perform detailed comparisons of synthetic infrared excesses with detected ones from Sun-like stars.  Applied to G2 stars, our models seek to demonstrate broad consistency while acknowledging the degeneracies involving $f_{\rm dust}$.  In other words, we simply wish to show that it is not difficult to produce synthetic infrared excesses around G2 stars that are of the same order of magnitude as in observed systems.

Figure \ref{fig:fir_hot} shows the infrared excess $f_{\rm IR}$ associated with dust grains produced in collisions between planetesimals residing in dynamically hot discs.  It is important to note that we are not intending for our calculations to be a rigorous prediction of detection statistics.  A few aspects of these results are worth emphasizing.  Consider the $f_{\rm dust}=1$ case.  Firstly, $f_{\rm IR}$ generally increases with the disc mass $M_{\rm disc}$ in a monotonic fashion.  Secondly, at a fixed value of $M_{\rm disc}$, there is a dispersion in the value of $f_{\rm IR}$ over several orders of magnitude.  This degeneracy between the infrared excess and disc mass arises from the fact that for a given value of $M_{\rm disc}$, there exists a large number of solutions of $r$ and $\sigma_r$ that satisfy the dynamical survival conditions.  Thus, $f_{\rm IR}$ is a poor diagnostic of the disc mass, as one expects.  Thirdly, the $t_{\rm coll} > t_\star$ condition sets an upper limit on the allowed value of $f_{\rm IR}$, because model discs with shorter collisions times associated with the \emph{planetesimals} are not expected to survive or persist for the stellar age.  For the 300 Myr-old M4 star, we have $f_{\rm IR} \lesssim 0.01$, consistent with the detection threshold stated in \cite{avenhaus12} for the WISE satellite.  By contrast, we have $f_{\rm IR} \lesssim 10$ for the G2 star, which is broadly consistent with the measured values of $f_{\rm IR} \sim 1$--10 associated with Sun-like stars \citep{bryden06,chen06,meyer08}.  Thus, our survival models simultaneously explain the null results of \cite{avenhaus12} involving M stars and are broadly consistent with the detections of previous studies examining Sun-like stars without any need for the finetuning of parameters.  Fourthly, we note that our suite of models suggest the existence of a vast population of planetesimal discs which are currently unobservable because the associated infrared emission is too faint to detect.  For comparison, the $a \approx 1$ AU planetesimal disc in HD 69830 is estimated to have a disc mass $M_{\rm disc} \lesssim 3$--$4 \times 10^{-3} M_\oplus$, several times more massive than our own asteroid belt \citep{heng11}.

When $f_{\rm dust} = 3.5$, the dust grains produced are either hotter than predicted by their blackbody temperatures at a given distance or emitting at a longer peak wavelength than the observed one.  Both interpretations require the grains to be situated farther away from the star by a factor of $f_{\rm dust}^2 = 12.25$.  Dynamical conditions become more forgiving and the region of allowed parameter space for the planetesimals is larger, as seen in Figure \ref{fig:rmdisc_hot}.  Corresponding, the infrared excesses allowed are typically higher as reflected in Figure \ref{fig:fir_hot}, because the allowed disc masses are generally higher.  If the sample of M stars examined by \cite{avenhaus12} have ages $\sim 300$ Myr, then a plausible, alternative interpretation is that they are somehow scrutinizing the fainter, undetectable members of the debris disc population with $f_{\rm dust} = 3.5$.

Figure \ref{fig:rmdisc_warm} shows the allowed planetesimal radius versus disc mass for a suite of dynamically warm discs.  The condition $t_{\rm coll} < t_\star$ plays the opposite role of $t_{\rm coll} > t_\star$: it sets an upper limit on the planetesimal radius for a given disc mass.  The maximum value of the planetesimal radius is set by a combination of demanding that the planetesimals are non-self-gravitating ($\Theta < 1$) and that the collisions between them are non-erosive, since both conditions depend on $\sigma_r$.  The non-erosive condition restricts the rest of the allowed parameter space.  The corresponding infrared excess from the \emph{planetesimals} is shown in Figure \ref{fig:fir_warm}.  The upper limit to $f_{\rm IR}$ for a wide range of $M_{\rm disc}$ is set by the minimum value of $r$ assumed in our Monte Carlo calculations, but this ultimately does not affect the maximum value of $f_{\rm IR}$ for a given suite of model discs.  Rather, it is set by the non-erosive condition, which depends on the value of $Q^\ast_{\rm D}$, the binding energy of the planetesimals.  The most uncertain quantity in the calculation of $\mbox{max}\{f_{\rm IR}\}$ is the normalization $Q_0$ in equation (\ref{eq:bind}) for the binding energy---the computed infrared excess scales roughly linearly with $Q_0$.  Higher values of $Q_0$ allow more planetesimals of a given size to be contained within a dynamically warm disc and still avoid dust-manufacturing collisions, thereby allowing for higher disc masses and infrared excesses.

Performing further calculations with $t_\star = 300$ Myr (not shown), we estimate that the predicted values of $f_{\rm IR}$ are consistent with the null results of \cite{avenhaus12} if $Q_0 \lesssim 10^4$ erg g$^{-1}$ and $f_{\rm dust}=1$.  For older dynamically warm discs, the range of $Q_0$ values allowed is less restrictive.  Generally, our model warm discs are also broadly consistent with the reported values of $f_{\rm IR}$ for Sun-like stars even allowing for $Q_0 \sim 10^4$--$10^7$ erg g$^{-1}$.  Overall, \emph{if an observed disc system is a dynamically warm planetesimal disc, then it tends to be more massive than ``traditional" debris discs (arising from dynamically hot discs) to compensate for the relative faintness of the infrared emission from the planetesimals.}

\begin{figure*}
\begin{center}
\includegraphics[width=\columnwidth]{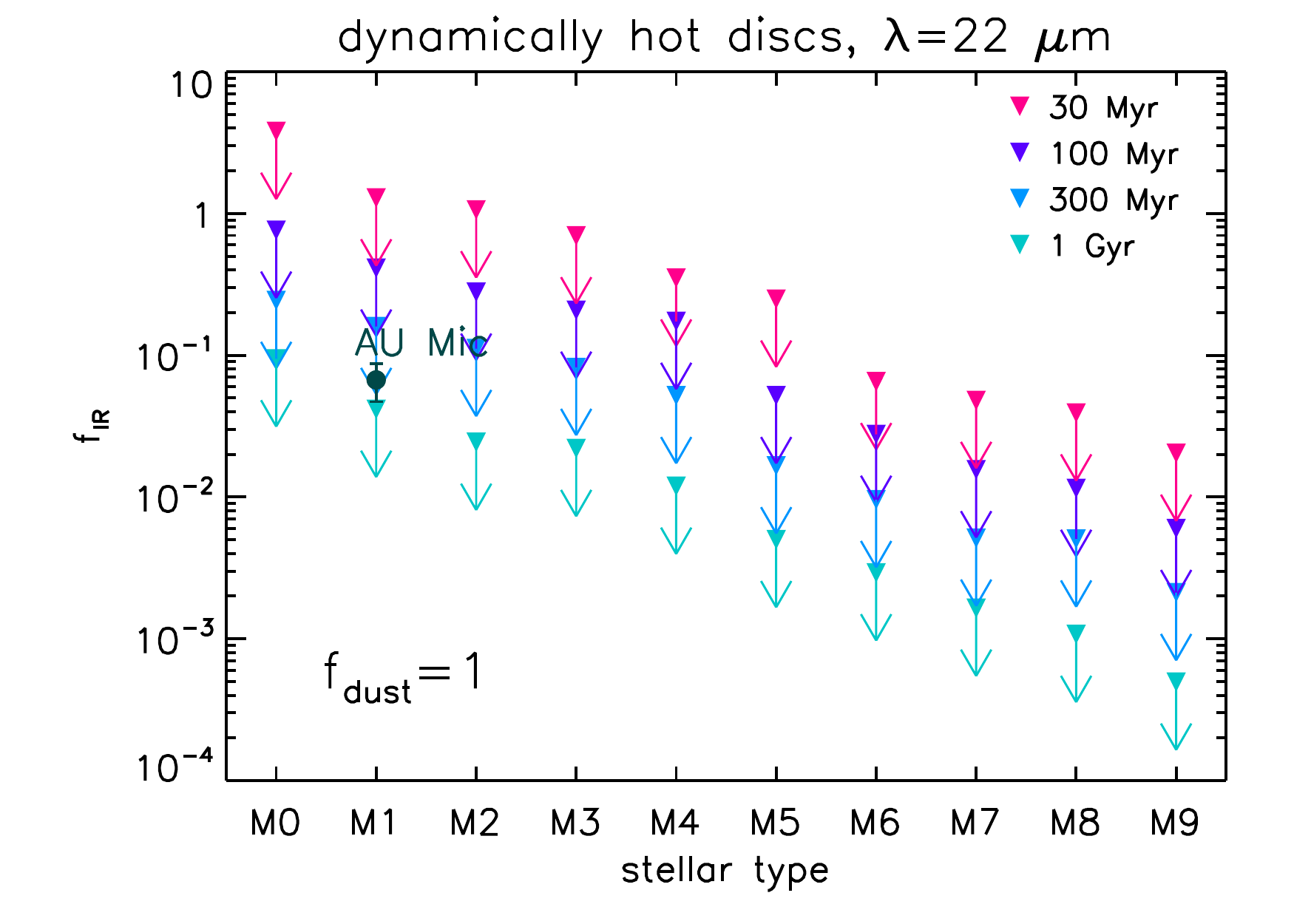}
\includegraphics[width=\columnwidth]{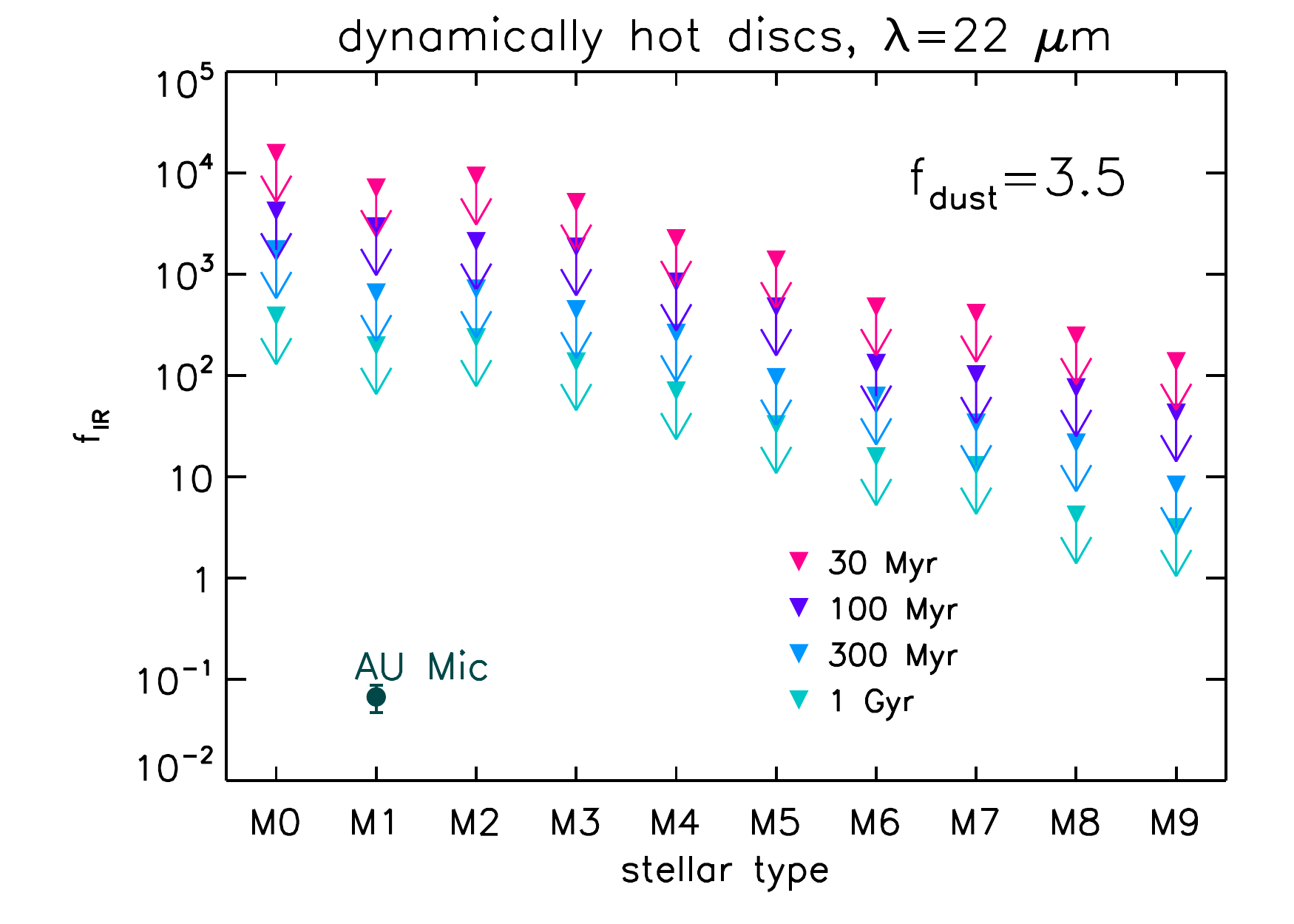}
\includegraphics[width=\columnwidth]{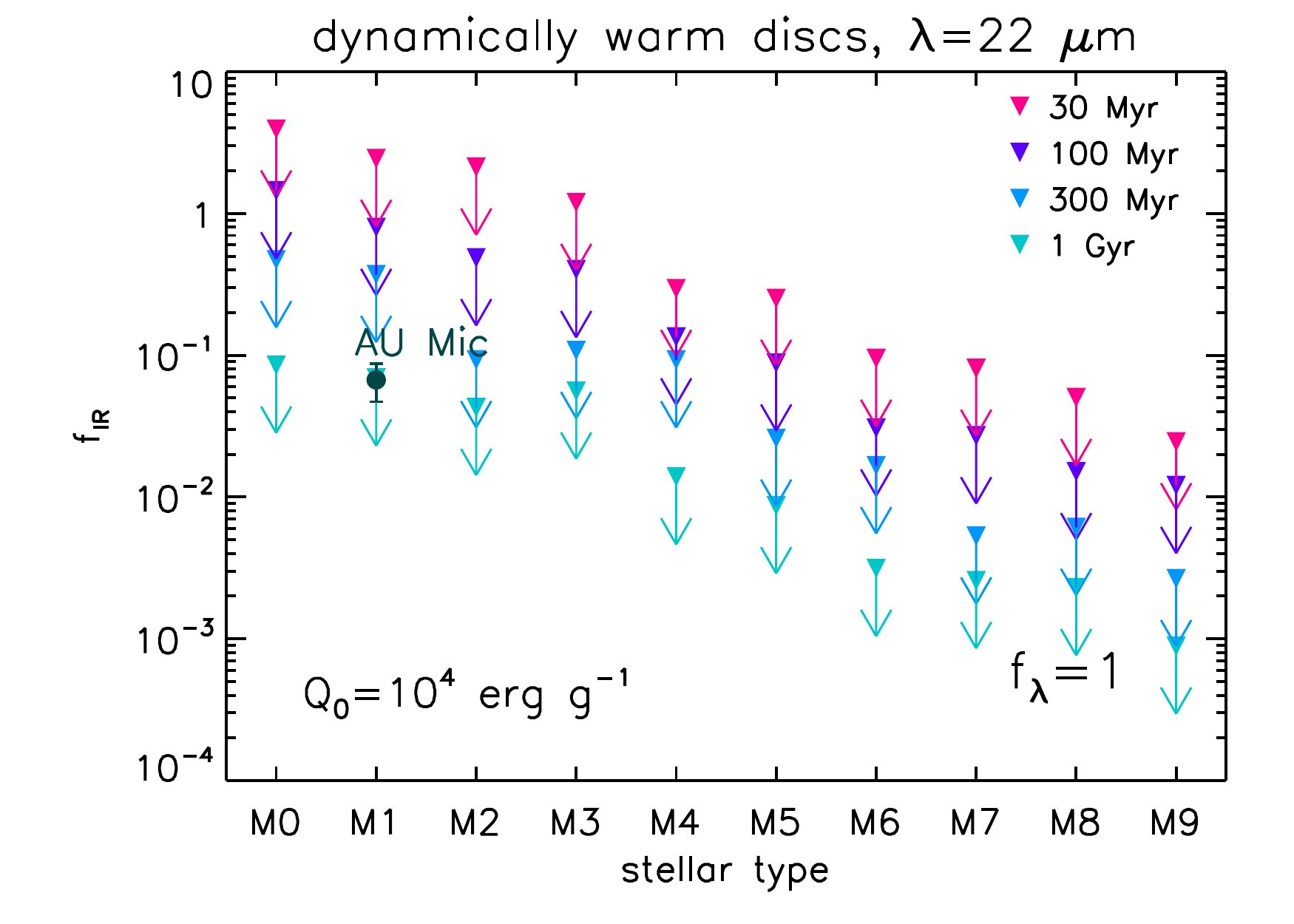}
\includegraphics[width=\columnwidth]{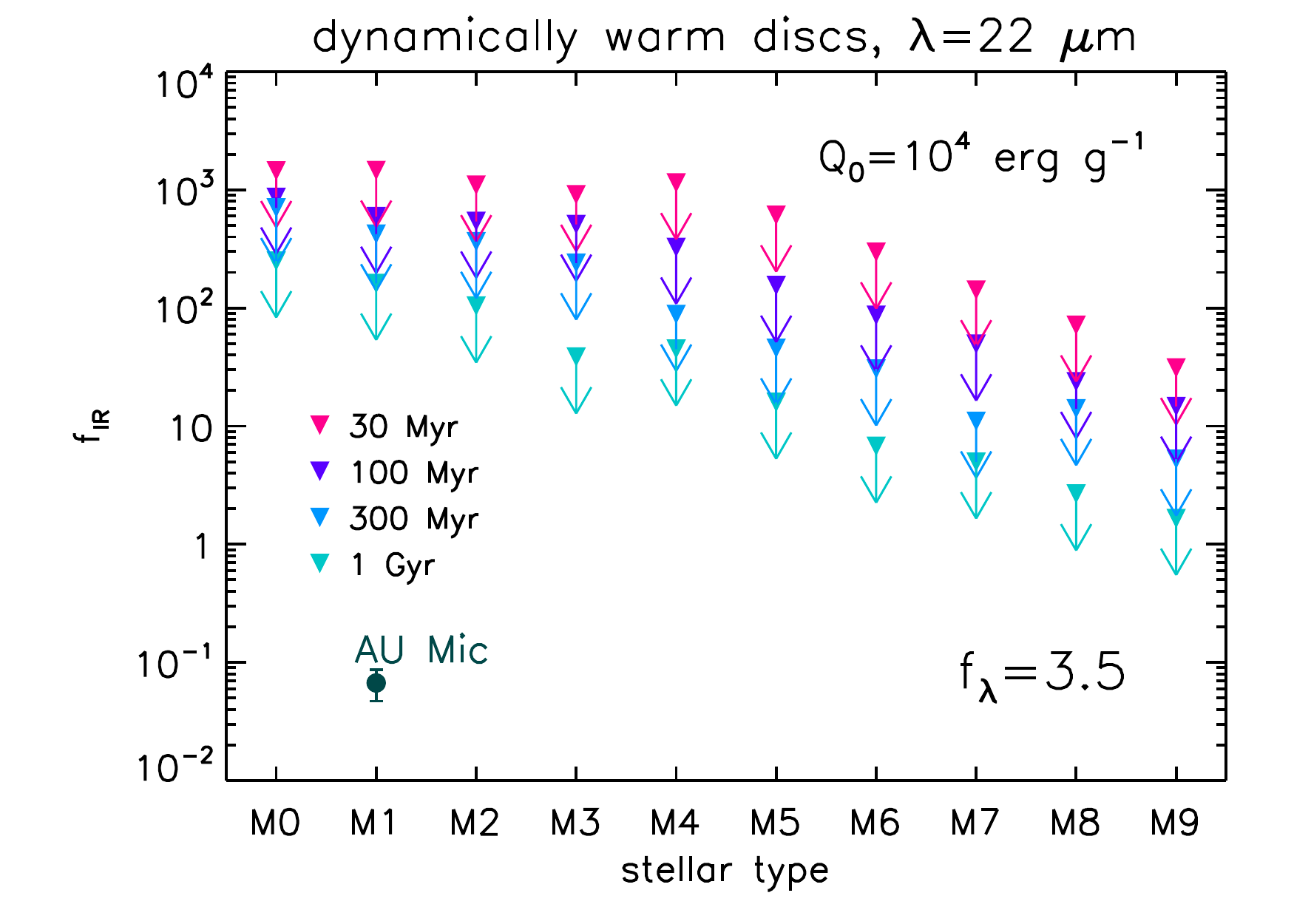}
\end{center}
\vspace{-0.2in}
\caption{Maximum infrared excesses (triangles) as computed by a Monte Carlo implementation of the survival models across the M spectral types.  Shown is the data point for AU Mic (circle; $t_\star \approx 12$ Myr), which is consistent with the maximum infrared excess allowed.  Top row: dynamically hot discs, which are the progenitors of debris discs in the traditional sense.  Bottom row: dynamically warm discs (assuming $Q_0=10^4$ erg g$^{-1}$).  Left column: $f_{\rm dust} =1$.  Right column: $f_{\rm dust} =3.5$.} 
\label{fig:firmax}
\end{figure*}

\subsection{Maximum Infrared Excess and Comparison to Observations}

The maximum infrared excess is computed across all M spectral types in Figure \ref{fig:firmax} for $t_\star = 30$, 100 and 300 Myr as well as 1 Gyr.  Dynamically warm discs produce similar values of $\mbox{max}\{f_{\rm IR}\}$ to dynamically hot discs if we assume $Q_0=10^4$ erg g$^{-1}$.  As expected, the maximum value of $f_{\rm IR}$ declines with both stellar age and luminosity, the latter arising from the effect of probing smaller values of $a$ as previously discussed.  

For comparison, we include the detection of an infrared excess from AU Mic at $\lambda=22$ $\mu$m by \cite{avenhaus12} (see \S\ref{sect:obs}).  The data point is easily consistent with the maximum infrared excess allowed for its age ($t_\star \approx 12$ Myr), whether we have $f_{\rm dust}=1$ or 3.5.  (The ``birth ring" of AU Mic is observed to reside at about 35 AU rather than 1 AU, cf. \citealt{wilner12}.) 

\begin{figure*}
\begin{center}
\includegraphics[width=\columnwidth]{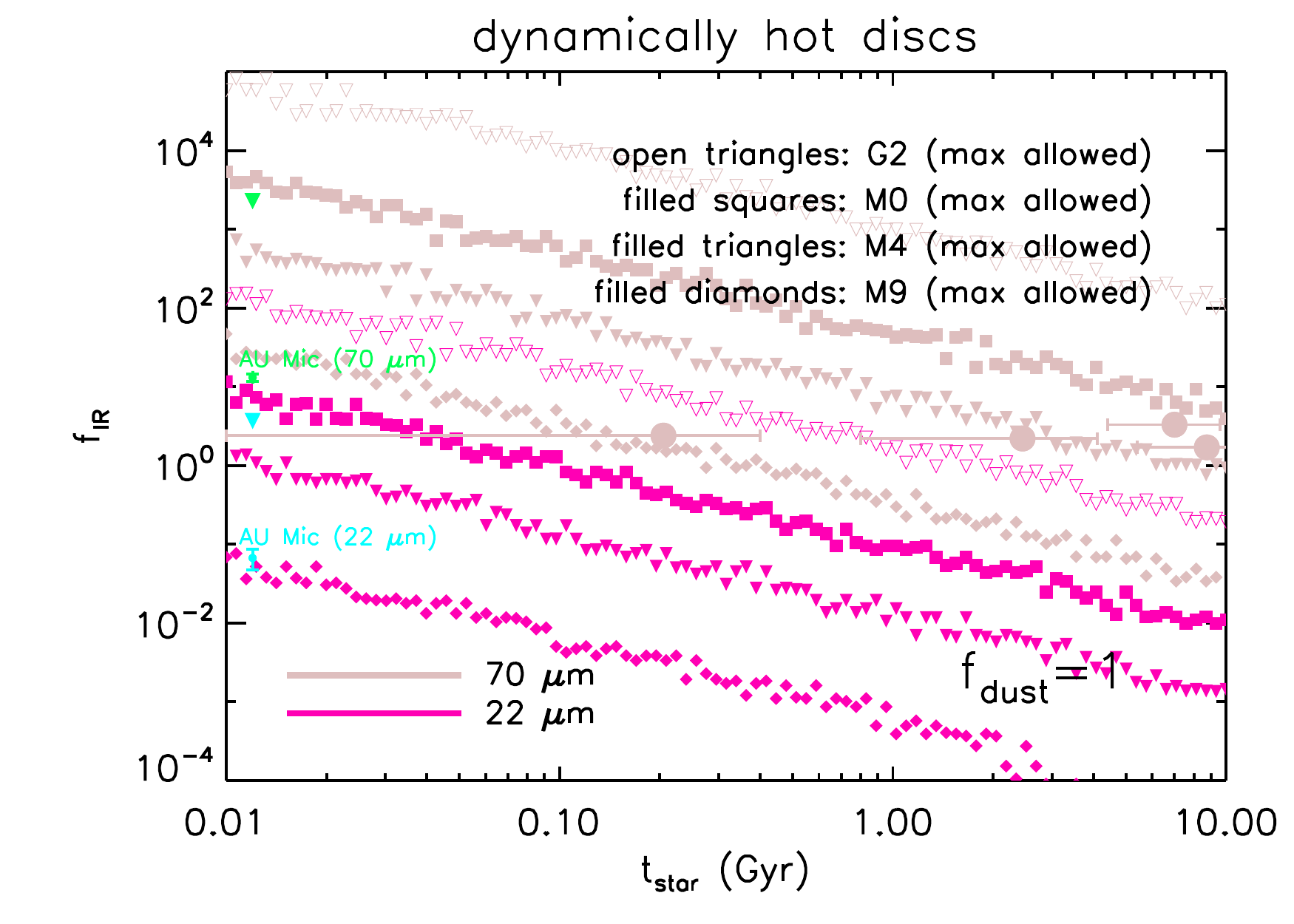}
\includegraphics[width=\columnwidth]{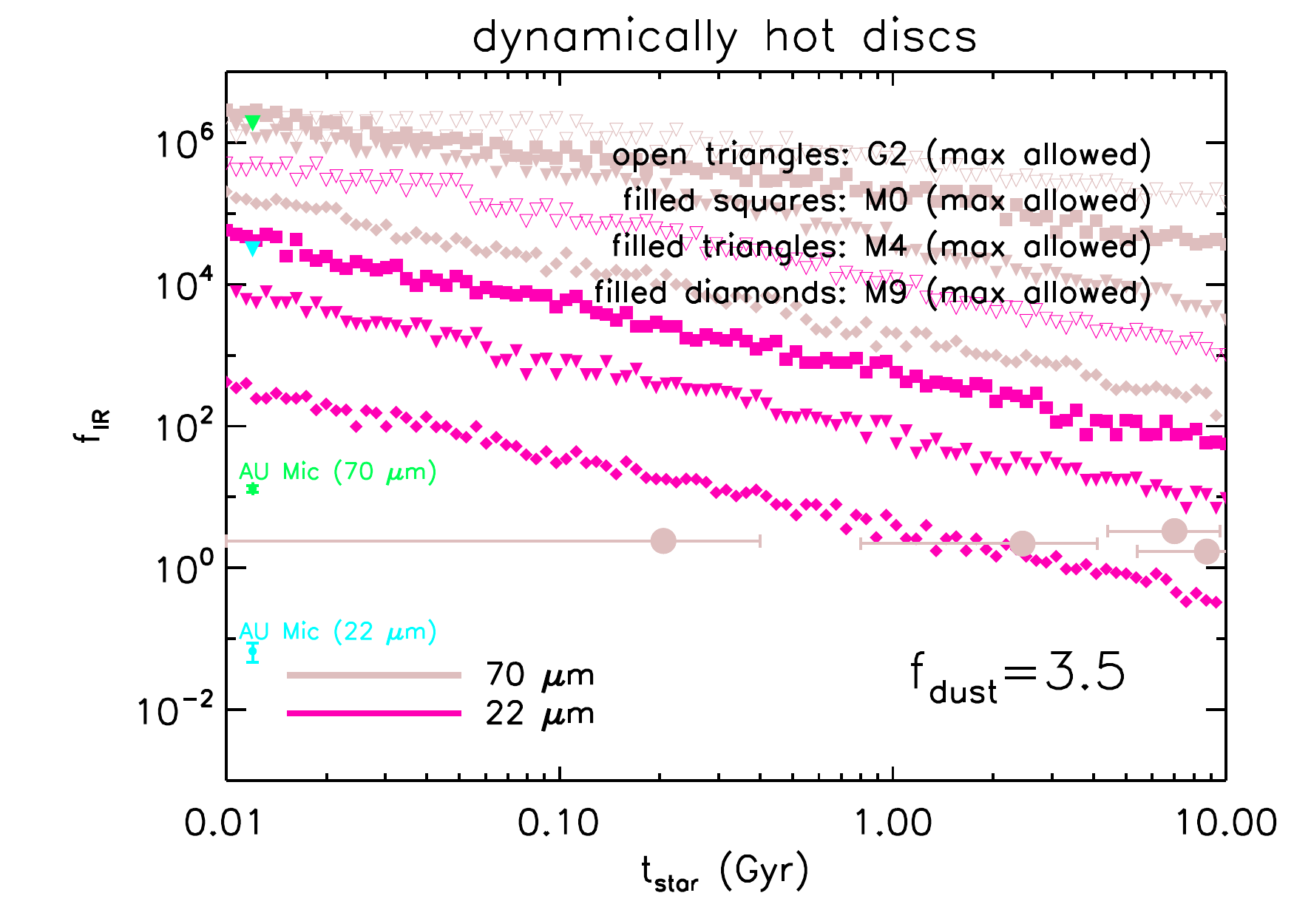}
\includegraphics[width=\columnwidth]{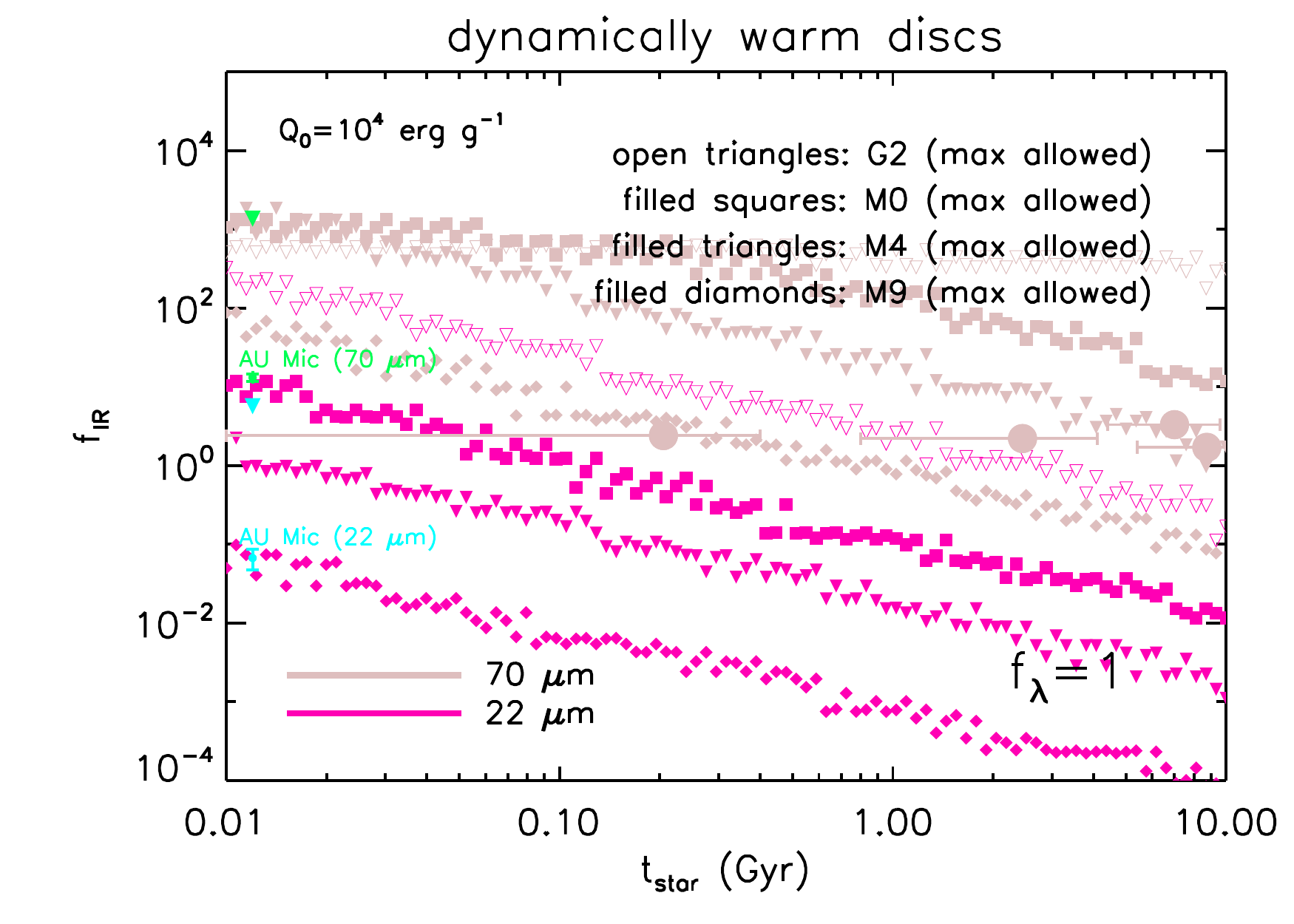}
\includegraphics[width=\columnwidth]{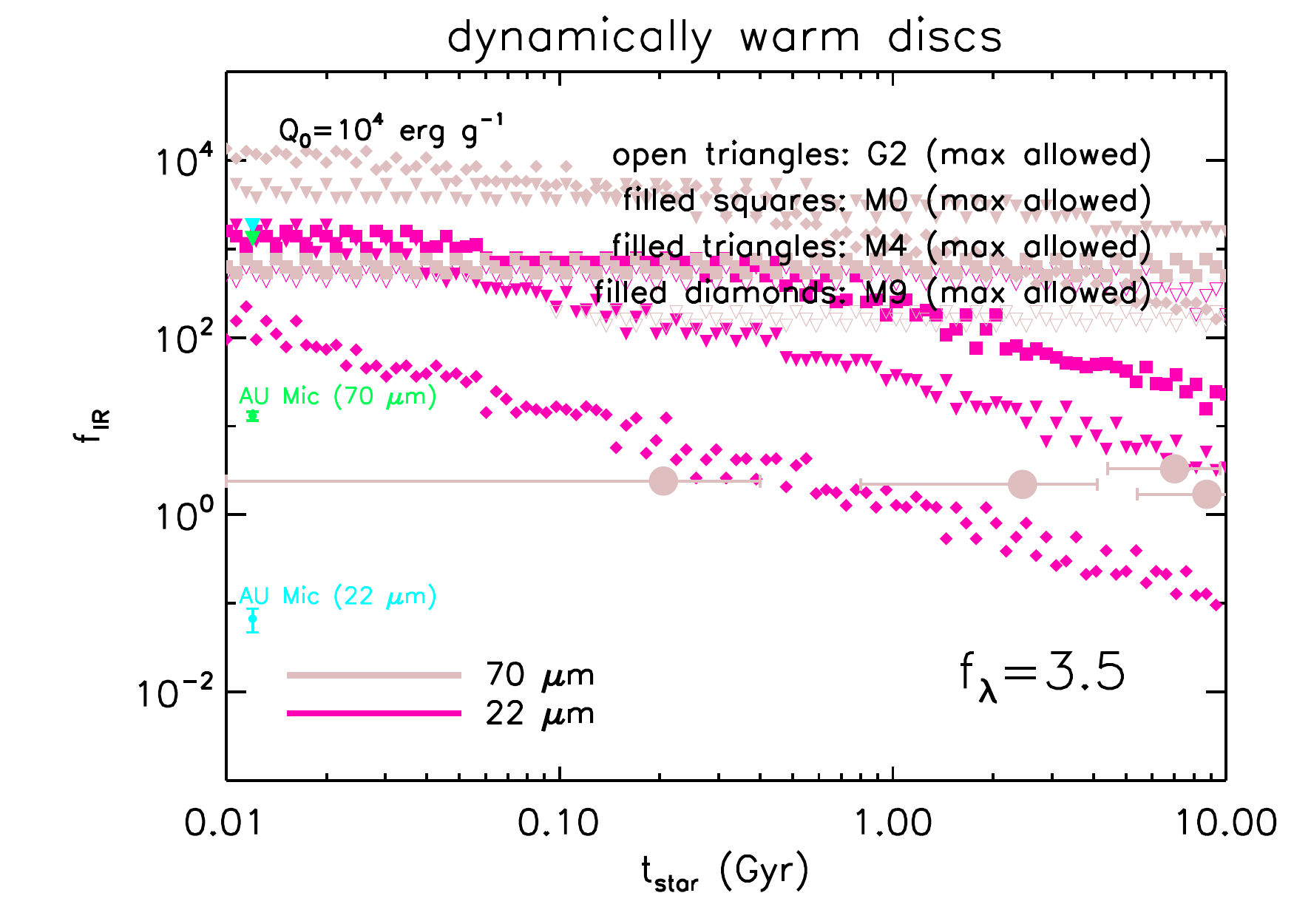}
\end{center}
\vspace{-0.2in}
\caption{Maximum infrared excesses (various symbols) as computed by a Monte Carlo implementation of the survival models across a range of stellar ages.  Shown are calculations for M0, M4, M9 and G2 stars at 22 $\mu$m and 70 $\mu$m.  For AU Mic, we assume a M1 star with $t_\star \approx 12$ Myr, compute the upper limits on the infrared excess (triangles) and compare them to the data points at 22 $\mu$m and 70 $\mu$m (circles).  Also shown are the measured $f_{\rm IR}$ at 70 $\mu$m from four G stars by Bryden et al. (2006); note that the quoted errors associated with $f_{\rm IR}$ are smaller than the diameters of the filled circles.  Top row: dynamically hot discs, which are the progenitors of debris discs in the traditional sense.  Bottom row: dynamically warm discs, where the saturation of $f_{\rm IR}$ for M0 and G2 stars at 70 $\mu$m arises from our assumption of a minimum radius of 1 cm for the planetesimals.  Left column: $f_{\rm dust} =1$.  Right column: $f_{\rm dust} =3.5$.}
\label{fig:firmax2}
\end{figure*}

Figure \ref{fig:firmax2} shows the maximum infrared excess across a range of stellar ages, $t_\star = 0.01$--10 Gyr.  Generally, we have $f_{\rm IR} \propto t_\star^{-1}$ in agreement with \cite{dd03}, \cite{wyatt07} and \cite{ht10} for dynamically hot discs.  Both dynamically hot and warm planetesimal discs around Sun-like stars generally produce brighter infrared emission due to the more favourable dynamical conditions present for the planetesimals to survive for long time scales.  We include in Figure \ref{fig:firmax2} the reported values of $f_{\rm IR}$ (at 70 $\mu$m) and $t_\star$ for G stars from \cite{bryden06}.  We do not specialize to the individual stellar parameters associated with these four objects, because doing so will only introduce minor corrections and does not change our overall result and conclusion.  Our computed upper limits on $f_{\rm IR}$ for G stars are easily consistent, by several orders of magnitude, with these measured values, implying the existence of model disc solutions capable of reproducing the measured $f_{\rm IR}$ values.  For $t_\star \gtrsim 300$ Myr, our models predict that debris discs around M stars are at best marginally detectable ($f_{\rm IR} \lesssim 0.1$) for spectral type M4 and later if $f_{\rm dust}=1$.  At smaller wavelengths (3.4 $\mu$m, 4.6 $\mu$m and 12 $\mu$m; not shown), our model discs produce lower values of the maximum allowed infrared excess.

We note that the $f_{\rm IR} \ggg 1$ values are implausible because the corresponding covering fraction associated with either the dust grains (for debris discs) or planetesimals (for dynamically warm discs) becomes optically thick.

Additionally, we compute the theoretical upper limits on $f_{\rm IR}$ associated with AU Mic.  It is apparent that these upper limits are again easily consistent with the detected infrared excesses at 22 $\mu$m and 70 $\mu$m (see \S\ref{sect:obs}).  We do not invert the measured infrared excesses from AU Mic to obtain an estimate for the disc mass, because it has been previously shown that at $\sim 10$ Myr the disc mass is essentially unconstrained \citep{heng11}.

Finally, we consider the recent discovery of an old debris disc around the M3 star in GJ 581 via the Herschel Space Observatory, using the PACS instrument, at 70, 100 and 160 $\mu$m by \cite{lestrade12}.  We use the stellar parameters listed in Table 2 of \cite{lestrade12} and take the average of the range of stellar ages quoted (5 Gyr).  At 70, 100 and 160 $\mu$m, we obtain $\mbox{max}\{f_{\rm IR}\} \approx 3$, 25 and 200, respectively, for dynamically hot discs if $f_{\rm dust}=1$.  The values of $\mbox{max}\{f_{\rm IR}\}$ become even larger for $f_{\rm dust}>1$.  For dynamically warm discs, we obtain $\mbox{max}\{f_{\rm IR}\} \approx 9$, 80 and 900 at the same respective wavelengths for $Q_0=10^4$ erg g$^{-1}$ and $f_{\rm dust}=1$; these upper limits are higher for larger values of $Q_0$ and $f_{\rm dust}$.  These computed values of the maximum infrared excesses are consistent with the values of about 3, 7 and 20 reported in \cite{lestrade12}.  (See their Figure 6.)

\section{Discussion}
\label{sect:discussion}

\subsection{Summary}

We have used survival models to examine the occurrence of planetesimal and debris discs around M and G stars.  The salient points of our investigation include:
\begin{itemize}

\item The dynamical survival conditions severely restrict the range of disc masses and planetesimal sizes allowed for a disc to persist for its stellar age.  Despite this, the infrared excess is a poor diagnostic of the disc mass at young ages ($\sim 300$ Myr), generally spanning several orders of magnitude for a given disc mass.\footnote{But see \cite{heng11} for an application of the survival models to estimate the disc mass when $t_\star \sim 1$ Gyr.}

\item The dearth of detectable infrared excesses from M stars, as probed by the WISE satellite, is possibly due to the small semi-major axes probed at 3.4--22 $\mu$m.  These small semi-major axes translate to less forgiving dynamical survival conditions for the planetesimal discs.  In other words, planetesimals may not exist at small distances from the star for time scales $\gtrsim 300$ Myr, regardless of whether the spectral type is M or G.  This interpretation becomes less clear when the dust grains (in debris discs) or planetesimals (in dynamically warm discs) deviate from blackbody behaviour (i.e., are hotter) \emph{or} emit predominantly at wavelengths longer than the observed one.  However, the interpretation is acceptable when they are hotter than blackbody \emph{and} emit at peak wavelengths shorter than the observed one.  Clearly, some of these degeneracies may be broken by searching for debris discs around M stars across a broad range of wavelengths.

\item Both dynamically hot and warm planetesimal discs are capable of producing infrared excesses that are consistent with the reported values for G stars.  It is plausible that both dynamically hot and warm planetesimal discs will generate detectable infrared excess around M stars as well, the former through dust grains generated during planetesimal collisions and the latter via the planetesimals reprocessing the starlight themselves.

\item The detected infrared excesses from AU Mic and GJ 581 are easily reconciled with youth and the large distances probed by the observations, respectively.  These conclusions are unaffected by the degeneracies involving the non-blackbody behaviour (or peak versus observed wavelength probed) of dust grains.

\end{itemize}

\subsection{Are Non-Gravitational Forces Important?}
\label{subsect:nongrav}

\begin{figure}
\begin{center}
\includegraphics[width=\columnwidth]{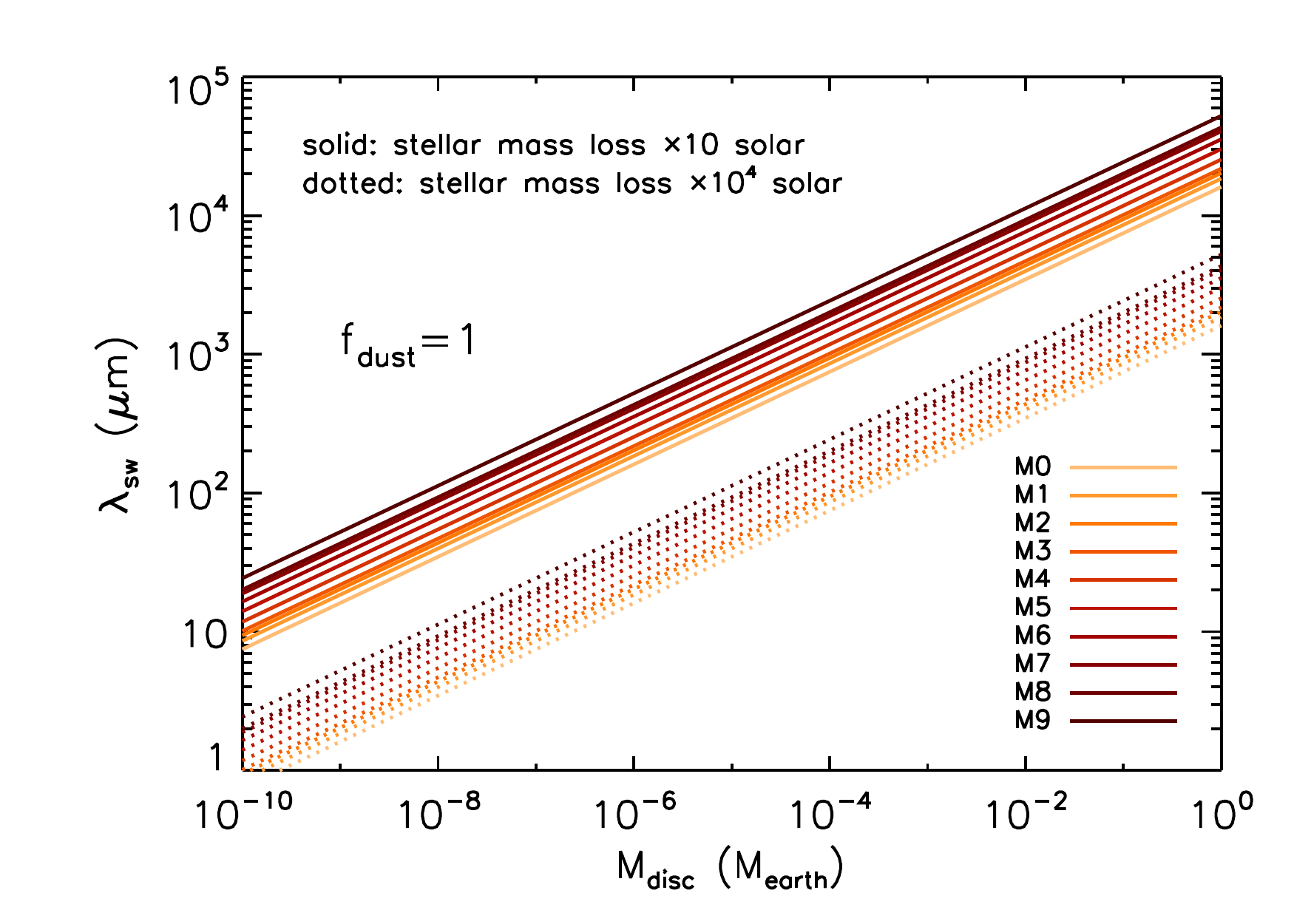}
\includegraphics[width=\columnwidth]{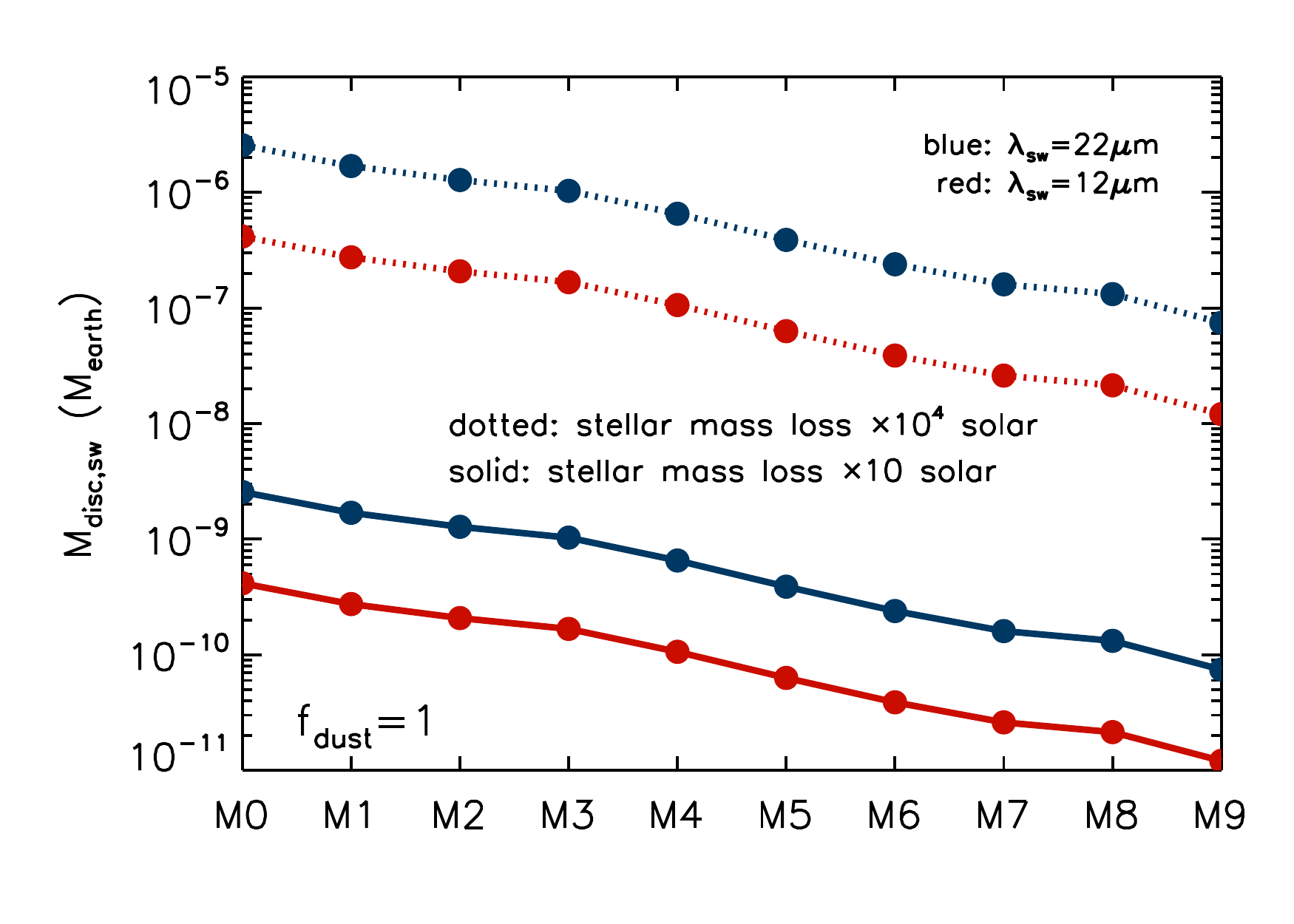}
\end{center}
\vspace{-0.2in}
\caption{Top panel: maximum observed wavelength below which stellar wind drag may be neglected from our analysis.  Bottom panel: minimum disc mass above which stellar wind drag may be neglected from our analysis.}
\label{fig:nongrav}
\end{figure}

A reasonable concern with our analysis is that we have neglected the effects of non-gravitational forces.  For Sun-like stars, \cite{ht10} previously demonstrated that Poynting-Robertson drag may be neglected when computing the emission properties of dust grains (see their \S5).  However, this may not generally be the case for M stars, since \cite{plavchan05} and \cite{sc06} have shown that stellar wind drag may dominate over Poynting-Robertson drag.  This result may be reproduced by examining the efficiency factor describing both types of non-gravitational drag,
\begin{equation}
Q_{\rm drag} = Q_{\rm PR} + \frac{\dot{M}_\star c^2}{{\cal L}_\star},
\end{equation}
where $Q_{\rm PR} \lesssim 1$ is the efficiency factor of Poynting-Robertson drag.  When $2 \pi r / \lambda \ll 1$, we have $Q_{\rm PR} \sim 2 \pi r / \lambda$.  We estimate that the stellar luminosity for M dwarfs is ${\cal L}_\star = 4 \pi R_\star^2 \sigma_{\rm SB} T^4_\star \sim 10^{-4}$--$10^{-1} {\cal L}_\odot$ with ${\cal L}_\odot$ denoting the solar luminosity.  The stellar mass loss rate for M stars is $\dot{M}_\star \sim 10$--$10^4 \dot{M}_\odot$ with $\dot{M}_\odot \approx 1.3 \times 10^{12}$ g s$^{-1}$ denoting the solar mass loss rate (see \citealt{plavchan05} and references therein).  Essentially, we infer that
\begin{equation}
\dot{M}_\star \gg \dot{M}_\odot \mbox{ and } {\cal L}_\star \ll {\cal L}_\odot \Rightarrow \frac{\dot{M}_\star c^2}{{\cal L}_\star} \gg Q_{\rm PR}.
\end{equation}
Thus, we have $Q_{\rm drag} \approx \dot{M}_\star c^2/{\cal L}_\star$ for M stars, consistent with the conclusions of \cite{plavchan05} and \cite{sc06}.

That stellar wind drag acts much faster than Poynting-Robertson drag, for M stars, does not address the issue of whether inter-grain collisions occur on an even faster time scale, in which case they are destroyed before experiencing significant orbital decay \citep{sc06}.  For Sun-like stars, \cite{wyatt05} and \cite{ht10} showed that collisions generally dominate over Poynting-Robertson drag.  We now elucidate the conditions under which collisions dominate over stellar wind drag for dust populations around M stars.

Denoting the radius of a dust grain by $r$, the time scale for Poynting-Robertson and stellar-wind drag to act on it is \citep{burns79}
\begin{equation}
t_{\rm drag} = \frac{8 \pi c^2 a^2 \rho r}{3 {\cal L}_\star Q_{\rm drag}} \approx \frac{8 \pi a^2 \rho r}{3 \dot{M}_\star}.
\end{equation}
The collisional time scale associated with the dust grains is \citep{ht10}
\begin{equation}
t_{\rm coll} = \frac{\pi^2 f_m \rho r}{12 f_1 M_{\rm disc}} \left( G M_\star \right)^{-1/2} a^{7/2}.
\end{equation}
Stellar wind drag may be neglected ($t_{\rm coll} < t_{\rm drag}$) when 
\begin{equation}
a < \left( \frac{32 f_1 M_{\rm disc}}{\pi f_m \dot{M}_\star} \right)^{2/3} \left( G M_\star \right)^{1/3}.
\label{eq:adrag}
\end{equation}

An alternative way of understanding equation (\ref{eq:adrag}) is to recast it using equation (\ref{eq:adust}), which yields an inequality for the observed wavelength,
\begin{equation}
\lambda < \lambda_{\rm sw},
\label{eq:lambda_sw}
\end{equation}
where
\begin{equation}
\lambda_{\rm sw} \equiv \frac{{\cal C}_{\rm Wien}}{T_\star f_{\rm dust}} \left( \frac{2}{R_\star} \right)^{1/2} \left( \frac{32 f_1 M_{\rm disc}}{\pi f_m \dot{M}_\star} \right)^{1/3} \left( G M_\star \right)^{1/6}.
\end{equation}
Generally, observing a debris disc at longer wavelengths corresponds to examining it at larger distances from the star.  Collisions occur on a time scale of $t_{\rm coll} \propto a^{7/2}$, while stellar wind drag occurs on a time scale of $t_{\rm drag} \propto a^2$, implying that there is a maximum distance beyond which collisions do not occur fast enough.  This can be compensated by having a more massive disc (higher $M_{\rm disc}$), a lower stellar mass loss rate (lower $\dot{M}_\star$) or a cooler star (lower $T_\star$).  All of these properties are reflected in equation (\ref{eq:lambda_sw}).

Yet another approach is to derive the minimum disc mass above which stellar wind drag may be neglected (i.e., $M_{\rm disc} > M_{\rm disc,sw}$),
\begin{equation}
M_{\rm disc,sw} = \frac{\pi f_m \dot{M}_\star}{32 f_1} \left( \frac{\lambda_{\rm sw} T_\star f_{\rm dust}}{{\cal C}_{\rm Wien}} \right)^3 \left( \frac{R_\star}{2} \right)^{3/2} \left( G M_\star \right)^{-1/2}.
\label{eq:mdisc_sw}
\end{equation}
Essentially, by estimating values for $M_{\rm disc,sw}$ we can evaluate the regions of parameter space, as shown in our figures, where stellar wind drag does not affect our computed results.

Figure \ref{fig:nongrav} shows calculations for $\lambda_{\rm sw}$ as a function of $M_{\rm disc}$ across the stellar types M0 to M9.  Given the large uncertainty in the stellar mass loss rate for M stars, we include calculations for both $\dot{M}_\star = 10 \dot{M}_\odot$ and $10^4 \dot{M}_\odot$.  It is apparent that for $M_{\rm disc} \gtrsim 10^{-4} M_\oplus$, we obtain $\lambda_{\rm sw} \sim 100$--1000 $\mu$m for $f_{\rm dust}=1$; for $f_{\rm dust}=3.5$, $\lambda_{\rm sw}$ decreases by a factor of $f_{\rm dust}$.  These estimates imply that for debris discs that are at least roughly as massive as our asteroid belt, the effects of stellar wind drag may be neglected when they are observed at infrared wavelengths.  More precise constraints may be obtained by examining the values of $M_{\rm disc,sw}$, also shown in Figure \ref{fig:nongrav}.  To relate these estimates to both the observations of \cite{plavchan05} and \cite{avenhaus12}, we adopt $\lambda_{\rm sw} = 12$ and 22 $\mu$m, respectively.  Generally, the minimum disc mass above which stellar wind drag may be neglected is low: $M_{\rm disc,sw} \sim 10^{-11}$--$10^{-6} f^3_{\rm dust} M_\oplus$.  \cite{plavchan05} searched for $\lambda=12$ $\mu$m infrared excesses around 9 M stars, but did not report any detections.  If stellar wind drag is to be invoked as an explanation for these non-detections, the discs around these M stars must have masses lower than the $M_{\rm disc,sw}$ values estimated at $\lambda_{\rm sw}=12$ $\mu$m (even leaving aside the issue that such low-mass discs probably emit infrared radiation below the detection limits).

Among the ensembles of both debris discs and dynamically warm planetesimal discs simulated by our Monte Carlo calculations, the most massive members are unaffected by stellar wind drag.  Even considering the wide range of stellar mass loss rates associated with M stars, the discs that are affected by stellar wind drag have low enough masses that they are undetectable using current instrumentation.  When stellar wind drag is considered for these low-mass discs, the associated dust grains (or planetesimals, if considering dynamically warm discs) become both undetectable and non-existent.

\subsection{Future Prospects}

Our results in Figures \ref{fig:firmax} and \ref{fig:firmax2} suggest that M stars with $t_\star \gtrsim 300$ Myr are at best marginally detectable by the WISE satellite at 22 $\mu$m, while those with $t_\star \sim 10$ Myr should be easily detectable---an age effect.  The caveat to this statement is that the dust grains (in debris discs) or planetesimals (in dynamically warm discs) either behave like or mimic blackbodies emitting predominantly at the observed wavelength (see \S\ref{subsect:ir} for details).  Conversely, if this caveat was strongly violated, then Figures \ref{fig:firmax} and \ref{fig:firmax2} predict that debris discs around M stars should be prevalent, a phenomenon that is not observed.  Considerations of formation, rather than survival, may further inform this issue.

The ability to compute $f_{\rm IR}$ values that are consistent with the measured ones for the 12 Myr-old AU Mic debris disc further buttresses the suggestion about the non-detections by WISE being an age effect.  Gas dispersal in protoplanetary discs around M stars is expected to take between 8 and 12 Myr \citep{simon12}, somewhat longer than for Sun-like stars.   \cite{schneider12} examined 30 K and M stars in the TW Hydrae association ($t_\star \approx 8$ Myr) and found that $42^{+10}_{-9}\%$ of them have infrared excesses of $f_{\rm IR} \sim 0.1$--10 (using the 22 $\mu$m channel of WISE), indicating the presence of dusty protoplanetary discs.  \cite{simon12} discovered a 6 Myr-old, pre-main-sequence, M4 star with $f_{\rm IR} \approx 40$ (again using the 22 $\mu$m channel of WISE) in the $\eta$ Cha star cluster.  (See also \citealt{gautier08}.)   No infrared excess was detected in the older star clusters: Tuc-Hor ($t_\star \approx 30$ Myr) and AB Dor ($t_\star \approx 70$ Myr).  Taken together, these results suggest that young ($t_\star \sim 10$ Myr) M stars should host debris discs that are detectable with WISE.

\vspace{0.2in}
\noindent
\textit{Acknowledgments}
KH acknowledges a tenure-track assistant professorship from the Center for Space and Habitability (CSH) at the University of Bern, an affiliate membership at the Institute for Theoretical Physics (ITP) at the University of Z\"{u}rich and partial financial support from the Swiss-based MERAC Foundation.  KH conducted part of the research while holding a Zwicky Prize Fellowship from ETH Z\"{u}rich in 2012.  We thank Michael Meyer, Hans Martin Schmid and Henning Avenhaus for interesting discussions.  We are grateful to the anonymous referee for constructive reports that improved the quality and clarity of the manuscript.


\label{lastpage}


\begin{thebibliography}{99}

\bibitem[Avenhaus, Schmid \& Meyer(2012)]{avenhaus12} Avenhaus, H., Schmid, H.M., \& Meyer, M.R. \ 2012, A\&A, 548, A105

\bibitem[Backman \& Paresce(1993)]{bp93} Backman, D.E., \& Paresce, F. \ 1993, in Protostars and Planets III, p. 1253--1304

\bibitem[Benz \& Asphaug(1999)]{ba99} Benz, W., \& Asphaug, E. \ 1999, Icarus, 142, 5

\bibitem[Bryden et al.(2006)]{bryden06} Bryden, G., et al. \ 2006, ApJ, 636, 1098

\bibitem[Burns et al.(1979)]{burns79} Burns, J.A., Lamy, P.L., \& Soter, S. \ 1979, Icarus, 40, 1

\bibitem[Chen et al.(2006)]{chen06} Chen, C.H., et al. \ 2006, ApJ, 166, 351

\bibitem[Dominik \& Decin(2003)]{dd03} Dominik, C., \& Decin, G. \ 2003, ApJ, 598, 626

\bibitem[Gautier et al.(2008)]{gautier08} Gautier III, T.N., Rebull, L.M., Stapelfeldt, K.R., \& Mainzer, A. \ 2008, ApJ, 683, 813

\bibitem[Heng \& Tremaine(2010)]{ht10} Heng, K., \& Tremaine, S. \ 2010, MNRAS, 401, 867

\bibitem[Heng(2011)]{heng11} Heng, K. \ 2011, MNRAS, 415, 3365

\bibitem[Kaltenegger \& Traub(2009)]{kt09} Kaltenegger, L., \& Traub, W.A. \ 2009, ApJ, 698, 519

\bibitem[Kenyon \& Bromley(2010)]{kb10} Kenyon, S.J., \& Bromley, B.C. \ 2010, ApJS, 188, 242

\bibitem[Krivov et al.(2008)]{krivov08} Krivov, A.V., M\"{u}ller, S., L\"{o}hne, T., \& Mutschke, H. \ 2008, ApJ, 687, 608

\bibitem[Lestrade et al.(2012)]{lestrade12} Lestrade, J.-F., et al. \ 2012, A\&A, 548, A86

\bibitem[Meyer et al.(2008)]{meyer08} Meyer, M.R., et al. \ 2008, ApJ, 673, L181

\bibitem[Plavchan et al.(2005)]{plavchan05} Plavchan, P., Jura, M., \& Lipscy, S.J. \ 2005, ApJ, 631, 1161

\bibitem[Plavchan et al.(2009)]{plavchan09} Plavchan, P., Werner, M.W., Chen, C.H., Stapelfeldt, K.R., Su, K.Y.L., Stauffer, J.R., \& Song, I. \ 2009, ApJ, 698, 1068

\bibitem[Rieke et al.(2005)]{rieke05} Rieke, G.H., et al. \ 2005, ApJ, 620, 1010

\bibitem[Rizzuto, Ireland \& Zucker(2012)]{rizzuto12} Rizzuto, A.C., Ireland, M.J., \& Zucker, D.B. \ 2012, MNRAS, 421, L97

\bibitem[Schneider, Melis \& Song(2012)]{schneider12} Schneider, A., Melis, C. \& Song, I. \ 2012, ApJ, 754, 39

\bibitem[Simon et al.(2012)]{simon12} Simon, M., Schlieder, J.E., Constantin, A.-M., \& Silverstein, M. \ 2012, ApJ, 751, 114

\bibitem[Stewart \& Leinhardt(2009)]{sl09} Stewart, S.T., \& Leinhardt, Z.M. \ 2009, ApJ, 691, L133

\bibitem[Strubbe \& Chiang(2006)]{sc06} Strubbe, L.E., \& Chiang, E.I. \ 2006, ApJ, 648, 652

\bibitem[Su et al.(2006)]{su06} Su, K.Y.L., et al. \ 2006, ApJ, 653, 675

\bibitem[Toomre(1964)]{toomre64} Toomre, A. \ 1964, ApJ, 139, 1217

\bibitem[Urban et al.(2012)]{urban12} Urban, L.E., Rieke, G., Su, K., \& Trilling, D.E. \ 2012, ApJ, 750, 98

\bibitem[Wilner et al.(2012)]{wilner12} Wilner, D.J., Andrews, S.M., MacGregor, M.A., \& Hughes, A.M. \ 2012, ApJL, 749, L27

\bibitem[Wyatt(2005)]{wyatt05} Wyatt, M.C. \ 2005, A\&A, 433, 1007

\bibitem[Wyatt et al.(2007)]{wyatt07} Wyatt, M.C., Smith, R., Su, K.Y.L., Rieke, G.H., \& Greaves, J.S. \ 2007, ApJ, 663, 365

\bibitem[Wyatt(2008)]{wyatt08} Wyatt, M.C. \ 2008, ARA\&A, 46, 339

\bibitem[Wyatt et al.(2012)]{wyatt12} Wyatt, M.C., et al. \ 2012, MNRAS, 424, 1206

\bibitem[Zuckerman(2001)]{z01} Zuckerman, B.\ 2001, ARA\&A, 39, 549

\end{thebibliography}
\end{document}